\documentclass[twocolumn,preprintnumbers,superscriptaddress]{revtex4-2}
\usepackage{amsmath}
\usepackage{amssymb}
\usepackage{mathrsfs}
\usepackage{graphicx}
\usepackage{bm}
\usepackage{epstopdf}
\usepackage{color}
\usepackage[colorlinks=true,citecolor=blue,linkcolor=blue,urlcolor=blue]{hyperref}
\usepackage{color}

\begin{document}
\title{Longitudinal and transverse mobilities of $n$-type monolayer transition
metal dichalcogenides in the presence of proximity-induced interactions at
low temperature}

\author{J. Liu}
\affiliation{School of Physics and Astronomy and Yunnan Key laboratory
of Quantum Information, Yunnan University, Kunming 650091, China}

\author{W. Xu}\email{wenxu$_$issp@aliyun.com}
\affiliation{School of Physics and Astronomy and Yunnan Key laboratory
of Quantum Information, Yunnan University, Kunming 650091, China}
\affiliation{Micro Optical Instruments Inc., Shenzhen 518118, China}
\affiliation{Key Laboratory of Materials Physics, Institute of Solid
State Physics, HFIPS, Chinese Academy of Sciences, HFIPS, Hefei 230031, China}

\author{Y. M. Xiao}\email{yiming.xiao@ynu.edu.cn}
\affiliation{School of Physics and Astronomy and Yunnan Key laboratory
of Quantum Information, Yunnan University, Kunming 650091, China}

\author{L. Ding}
\affiliation{School of Physics and Astronomy and Yunnan Key laboratory
of Quantum Information, Yunnan University, Kunming 650091, China}

\author{H. W. Li}
\affiliation{Micro Optical Instruments Inc., Shenzhen 518118, China}

\author{B. Van Duppen}
\affiliation{Department of Physics, University of Antwerp,
Groenenborgerlaan 171, B-2020
Antwerpen, Belgium}

\author{M. V. Milo\v{s}evi\'{c}}
\affiliation{Department of Physics, University of Antwerp,
Groenenborgerlaan 171, B-2020 Antwerpen, Belgium}

\author{F. M. Peeters}
\affiliation{Micro Optical Instruments Inc., Shenzhen 518118, China}
\affiliation{Department of Physics, University of Antwerp,
Groenenborgerlaan 171, B-2020 Antwerpen, Belgium}

\date{\today}

\begin{abstract}
We present a detailed theoretical investigation on the electronic transport
properties of $n$-type monolayer (ML) transition metal dichalcogenides
(TMDs) at low temperature in the presence of proximity-induced interactions
such as Rashba spin-orbit coupling (RSOC) and the exchange interaction.
The electronic band structure is calculated by solving the Schr\"{o}dinger
equation with a $\mathbf{k}\cdot\mathbf{p}$ Hamiltonian, and the electric screening
induced by electron-electron interaction is evaluated under a standard random phase approximation
approach. In particular, the longitudinal and transverse or Hall mobilities
are calculated by using a momentum-balance equation derived from a semi-classical
Boltzmann equation, where the electron-impurity interaction is considered as
the principal scattering center at low temperature. The obtained results
show that the RSOC can induce the in-plane spin components for spin-split
subbands in different valleys, while the exchange interaction can lift the
energy degeneracy for electrons in different valleys. The opposite signs
of Berry curvatures in the two valleys would introduce opposite directions
of Lorentz force on valley electrons. As a result, the transverse currents from
nondegenerate valleys can no longer be canceled out so that the transverse current
or Hall mobility can be observed. Interestingly, we find that at a fixed effective
Zeeman field, the lowest spin-split conduction subband in ML-TMDs can be tuned
from one in the $K'$-valley to one in the $K$-valley by varying the Rashba parameter.
The occupation of electrons in different valleys also varies with changing carrier
density. Therefore, we can change the magnitude and direction of the Hall
current by varying the Rashba parameter, effective Zeeman field, and carrier
density by, e.g., the presence of a  ferromagnetic substrate and/or applying a gate
voltage. By taking the ML-MoS$_2$ as an example, these effects are demonstrated
and examined. The important and interesting theoretical findings can be beneficial
to experimental observation of the valleytronic effect and to gaining an in-depth
understanding of the ML-TMDs systems in the presence of proximity-induced
interactions.
\end{abstract}
\maketitle

\section{Introduction}
In recent years, the investigation of transition-metal dichalcogenides (TMDs)
-based atomically thin two-dimensional (2D) electronic systems has attracted
a great deal of attention in condensed-matter physics and nano-electronics communities
due to their spintronic and valleytronic properties \cite{Xiao12,Cao12}.
These unique and interesting electronic properties are promising for advanced
electronics and optoelectronics, with potential applications in next-generation
information technology \cite{Mak18,Schaibley16,Mak16}. The discovery of 2D TMDs
-based valleytronic systems has also led to the proposal and observation of novel
physics effects such as the valley Hall effect (VHE) \cite{Xiao12,Bleu17-1,Bleu17-2}, which
is electrically equivalent to the Hall effect observed in the presence of an external
perpendicular magnetic field. One of the most interesting features of a free-standing
valleytronic material is that the electron energies are degenerated around $K$ and $K'$
points in the electronic band structure \cite{Dery15,Wang16,Liu13}. However, the electronic
spin orientations around these two valleys are just the opposite \cite{Xu14}. Thus, the
electronic band structure exhibits Berry curvature \cite{Xiao12,Xiao07}, and the
electrons with different spin orientations can move along different directions under
the action of a driving electric field and/or a polarized electromagnetic (EM)
field \cite{Xiao10,Lensky15}. Hence, the experimental techniques for the measurement
of the VHE are similar to those used for the detection of the spin Hall effect in
spintronic systems in the absence of an external magnetic field \cite{Kato04}.

The VHE has been observed experimentally in 2D TMDs systems normally in the presence
of driving EM field. This enables the electrical and optical detections and
manipulation of the photo-induced valley current in valleytronic systems \cite{Mak14}.
However, at present the experimental observation of the VHE by directly using the
electric transport measurement has been unsuccessful. There is also the lack of
related theoretical investigation to address this problem. From a viewpoint
of condensed matter physics, in an inversion symmetric breaking 2D ML-TMDs system with
energy degeneracy around $K$- and $K'$-point, under the action of a driving
electric field the current densities induced by electron movement from two valleys
are equal in magnitude but with opposite directions due to different signs of Berry
curvatures. As a result, the overall transverse or Hall voltage is canceled out
and the VHE cannot be observed by conventional transport measurement. A way out
of this situation is by breaking or lifting the valley degeneracy. One of the most
efficient schemes to lift the valley degeneracy in a 2D TMDs material is to
place the ML-TMDs film on a dielectric or ferromagnetic
substrate \cite{Qi15,Zhao17}. In such a case, the proximity effect induced
by the presence of the substrate can result in not only the Rashba spin-orbit
coupling (RSOC) \cite{Yang17,Soumyanarayanan16,Yao17} but also the exchange
interaction with an effective Zeeman field (EZF) \cite{Liang17}. The RSOC can
induce the in-plane electronic spin and the corresponding modification of
spin splitting in the electronic band structure in a ML-TMDs. This can be utilized for
the access and manipulation of valleys and spins in a 2D ML-TMDs/substrate
system. Furthermore, the proximity induced exchange interaction can lift
the valley degeneracy in the electronic energy spectra due to the introduction
of a magnetic momentum and/or van der Waals force in a 2D ML-TMDs system.
When the energy degeneracy in different valleys is lifted, the valley currents
from different valleys with different orientations can no longer be canceled
out. Thus, the electric voltage can be measured in transverse direction and,
hence, the VHE can be observed under the action of a dc electric field.
Consequently, in the presence of a particular substrate the proximity induced
interactions can be utilized for directly observation of the VHE in 2D ML-TMDs
systems using conventional transport measurement, namely applying an electric
current/voltage along the $x$-direction and measuring the voltage/current along
the $y$-direction. When the resistance $R_{xy}$ or conductance $\sigma_{xy}$
is nonzero, the electric VHE is observed.

Very recently, we have constructed an electron Hamiltonian in which the
RSOC and the EZF induced by proximity effects are considered for ML-TMDs
under the standard ${\bf k}\cdot {\bf p}$ approximation \cite{Zhao20}.
It has been demonstrated that in the presence of RSOC and EZF, the electronic
band structure in ML-TMDs depends strongly on the proximity-induced interactions
and the optical Hall effect can be observed by applying the linearly polarized
EM radiation field on a ML-TMDs \cite{Zhao20}, where a non-zero optical
conductivity $\sigma_{xy}(\omega)$ can be measured and the sign of the optical
Hall current or polarization can be tuned by varying the Rashba parameter. In
this study, we evaluate the longitudinal and transverse mobilities in a ML-TMDs
system by including the proximity-induced RSOC and exchange interaction. We take
the ML-MoS$_{2}$ as an example to examine the dependence of the electronic
screening and the longitudinal and transverse mobilities upon the strengthes of
the RSOC and the EZF. The prime motivation of this study is to see under what
conditions the VHE can be observed by direct electric transport measurement
and whether this effect is experimentally measurable. The paper is organized
as follows. The theoretical approaches developed in this study are presented in
Sec. \ref{sec:theoretical approach}, where we derive the formulas for the
calculations of the electronic band structure, Berry curvature, the inverse
electronic screening length, and the longitudinal and transverse or Hall
mobilities. The obtained results are presented and discussed in
Sec. \ref{sec:results and discussions} and the concluding remarks are summarized
in Sec. \ref{sec:conclusions}.

\section{Theoretical approach}
\label{sec:theoretical approach}
\subsection{Electronic band structure and Berry curvature}
In this study, we consider a ML-TMDs placed on a substrate with which the
proximity-induced interactions can lead to the enhancements of the valley
splitting and to Rashba spin-orbit coupling (SOC). We take an effective low-
electron energy Hamiltonian that includes the effect of RSOC and EZF induced
by the proximity interactions. The Hamiltonian for an electron in a ML-TMDs/substrate
heterostructure system consists of four parts\cite{Yao17,Zhao20,Liu23}:
\begin{align}\label{hamiltonian1}
H=H_0+H_{\text{SOC}}+H_{\text{ex}}+H_R,
\end{align}
where $H_0$ originates from the electronic orbital interaction, $H_{\text{SOC}}$ is the intrinsic
SOC in the system, $H_{\text{ex}}$ is attributable to the exchange interaction, and $H_{R}$
is the contribution from the RSOC. As we know, ML-TMDs is a 2D hexagonal crystal with
uniaxial symmetry where the RSOC always exists \cite{Min06,Eremeev12}. When the ML-TMDs is
placed on a substrate, the presence of the heterostructure can lead to the breaking of
the inversion symmetry along the direction normal to the 2D plane of the ML-TMDs \cite{Yao17,Liang17}.
This can further enhance the RSOC \cite{Dong08,Marchenko12,King11}.
Furthermore, the presence of the dielectric and/or ferromagnetic substrate can result
in an exchange interaction between the ML-TMDs and the substrate. The exchange interaction
is due to the introduction of EZF by the van der Waals force in the film/substrate
heterostructure \cite{Liang17} and adds the $H_{\text{ex}}$ term into Eq. \eqref{hamiltonian1}.
Meanwhile, the above Hamiltonian can be written in the form of a $4\times 4$ matrix given
as \cite{Zhao20}
\begin{equation}\label{hamiltonian12}
\setlength{\arraycolsep}{0.6pt}
\begin{split}
& H^\zeta({\mathbf k})={1\over 2} \times \\
& \left[\begin{array}{cccc}
\Delta+d_\zeta^{c}\ \ & 2A k_\zeta^- \ \ & 0 \ \ & i(1-\zeta )\lambda_R \\
2A k_\zeta^+ \ \ & d_\zeta^{v}-\Delta \ \ & -i(1+\zeta)\lambda_R & \ \ 0 \\
0\ \ & i(1+\zeta)\lambda_R\ \ & \Delta -d_\zeta^{c} & \ \ 2A k_\zeta^- \\
i(\zeta-1)\lambda_R \ \ & 0 \ \ & 2A k_\zeta^+ \ \ &
-(\Delta+d_\zeta^{v})
\end{array}\right],
  \end{split}
\end{equation}
where $\zeta =\pm$ refers to the $K$ ($K'$) valley, ${\bf k} = (k_x,k_y )$ is the
electron wave vector along the 2D-plane, $k_\zeta^\pm=\zeta k_x\pm ik_y$, $\Delta$
is the direct band gap between the conduction and valence bands, and $A=at$, with
$a$ being the lattice parameter and $t$ the hopping parameter \cite{Xiao12,Lu13,Li12}.
Furthermore, $d_\zeta^{\beta}=\zeta\lambda_\beta- B_\beta$, where $\beta=(c,v)$ refers to the
conduction and valence band, respectively. The intrinsic SOC parameter $2\lambda_c$
is the spin splitting at the bottom of the conduction band, and $2\lambda_v$ is that
at the top of the valence band in the absence of the RSOC \cite{Xiao12,Zhu11}.
$B_c$ and $B_v$ are the EZF experienced by an electron in the conduction and valence
bands in the presence of exchange interaction. $\lambda_R =\alpha_R\Delta/(2at)$ comes
from the RSOC, with $\alpha_R$ being the Rashba parameter \cite{Kormanyos14,Slobodeniuk16}.
The corresponding Schr\"odinger equation for a carrier in a ML-TMDs system near the
$K$ ($K'$) valley can be solved analytically. The four eigenvalues, $E=E_{\beta,s}^\zeta ({\bf k})$,
with $s=\pm $ being the spin index, are the solutions of the diagonalized equation of
the matrix, which reads
\begin{align}\label{diagonalized}
E^4-A_2E^2+A_1E+A_0=0,
\end{align}
with
\begin{align}
A_2=&\Delta^2/2 +\lambda_R^2
+2A^2k^2+({d_\zeta^v}^2 +{d_\zeta^c}^2)/4, \nonumber\\
A_1=&\Delta({d_\zeta^v}^2-{d_\zeta^c}^2)/4-\zeta\lambda_R^2
(d_\zeta^v-d_\zeta^c)/2,\nonumber\\
A_0=&\bigl({\Delta^2/4}+A^2k^2 \bigr)^2+\lambda_R^2(\Delta+\zeta d_\zeta^c)
(\Delta+\zeta d_\zeta^d)/4\nonumber\\
&-\Delta^2({d_\zeta^c}^2 +{d_\zeta^v}^2)/16-A^2k^2
d_\zeta^cd_\zeta^v/2+(d_\zeta^cd_\zeta^v)^2/16. \nonumber
\end{align}
The corresponding eigenfunctions for an electronic state near the $K$ and $K'$ points are
\begin{equation}\label{eigenfunctions}
|{\bf k};\lambda>={\cal A}_{\beta,s}^\zeta[c_1,c_2,c_3,c_4]e^{i{\bf k}\cdot{\bf r}},
\end{equation}
where ${\bf r}=(x,y)$, $\lambda=(\beta,\zeta,s)$,
\begin{align}
&c_1=i\lambda_R [h_1+4A^2 (1+\zeta)({k_\zeta^-})^2], \nonumber\\
&c_2=-4iA\lambda_Rk_\zeta^- h_2, \nonumber\\
&c_3=2Ak_\zeta^- h_3, \nonumber\\
&c_4=-[(\Delta-2E)^2 -({d_\zeta^c})^2)](\Delta+2E-{d_\zeta^v})\nonumber\\
    &\ \ \ \ \ \ -(1+\zeta)^2\lambda_R^2(\Delta-2E+d_\zeta^c)\nonumber\\
    &\ \ \ \ \ \ -4A^2k^2(\Delta-2E-d_\zeta^c), \nonumber
\end{align}
and
${\cal A}_{\beta,s}^\zeta ({\bf k})=(|c_1|^2+|c_2|^2+|c_3|^2+|c_4|^2)^{-1/2}$
is the normalization coefficient. Here,
$h_1=(1-\zeta)(\Delta -2E-d_\zeta^c) (\Delta+2E-d_\zeta^v)$,
$h_2=\Delta-2E+\zeta d_\zeta^c$, and
$h_3=(\Delta-2E+d_\zeta^c)(\Delta+2E-d_\zeta^v)+4A^2 k^2$.
As we know, the Rashba term $H_{R}$ can lead to the projection of the spin
component in the $x$-$y$ plane and mixes the spin states. Therefore, the spin
index $s$ is no longer a good quantum number \cite{Yao17,Ochoa13}, and thus
we use $s=\pm$ for up/down for the sake of distinguishing different
electronic states induced by RSOC.

As we known, the Berry curvature can significantly modify the electron dynamics
and generate new electrical transport phenomena by introducing an effective
magnetic field. For a free-standing ML-TMDs, the electrons in the two valleys
experience effective magnetic fields proportional to the Berry curvatures with
equal magnitudes but opposite signs due to the broken inversion symmetry in its
crystal structure \cite{Xiao12,Xiao10,Mak14}. However, the presence of proximity-induced
interactions such as RSOC and exchange interaction would also modify the Berry
curvature behaviors. With eigenvalues and eigenfunctions obtained from
Eqs.\eqref{diagonalized} and \eqref{eigenfunctions}, the Berry curvature of
ML-TMDs in the presence of the proximity-induced interactions
${\bf \Omega}_{\lambda}(\mathbf{k})=\nabla_{\mathbf{k}}\times i\langle{\mathbf{k}};\lambda|
\nabla_{\mathbf{k}}|{\mathbf{k}};\lambda\rangle\cdot\hat{\mathbf{z}}$
at each valley can be calculated through \cite{Kim18}
\begin{align}\label{berry}
{\bf \Omega}^\zeta_{\beta,s}(\mathbf{k})=&i\sum^{\prime}_{\beta',s'}
\bigg[\frac{\langle{\bf k};\lambda|
\partial H/\partial k_{x}|{\bf k};\lambda'\rangle\langle{{\bf k};\lambda'}|
\partial H/\partial k_{y}|{\bf k};\lambda\rangle}{[E^\zeta_{\beta,s}({\bf k})
-E^\zeta_{\beta',s'}({\bf k})]^2}\nonumber\\
&-\bigg(\frac{\partial}{\partial k_x}
\leftrightarrow\frac{\partial}{\partial k_y}\bigg)\bigg],
\end{align}
where the prime symbol ($'$) above $\sum$ is introduced to denote the exclusion
of the case $(\beta',s')=(\beta,s)$.

\subsection{Electron-electron interaction and electronic screening}
With the electron wave-function and the energy spectrum, we can
evaluate the electrostatic energy induced by electron-electron
(e-e) interaction and the dynamical dielectric function under the
usual random-phase approximation (RPA). From now on, we consider an
$n$-type ML-TMDs system in which the conducting carriers are electrons
in the conduction band. Taking only the spin splitting
conduction band (i.e., $\beta=c$ only) into account, we use the
notations $\psi_{s{\bf k}}^\zeta ({\bf r})$ and $E_{s}^\zeta({\bf
k})$ for the electron wave function and the energy spectrum, respectively,
at a state $|{\bf k};\zeta,s\rangle$ in the conduction band. As a result,
we are now dealing with a two-band situation around the $K$ or $K'$ points.
The electrostatic potential induced by the bare e-e interaction via the Coulomb
interaction $V({\bf r})=e^2/(\kappa |{\bf r}|)$ can be calculated via
\begin{align}
V_{s's}^\zeta({\bf k},{\bf q})=&\langle\psi_{s'{\bf k_1+q}}^{\zeta*} ({\bf
r}_1)\psi_{s{\bf k_1}}^{\zeta} ({\bf r}_1) |V({\bf r}_1-{\bf
r}_2)\nonumber\\
&|\psi_{s{\bf k_2}}^{\zeta*} ({\bf r}_2)\psi_{s'{\bf k_2+q}}^{\zeta}
({\bf r}_2)\rangle=V_q F_{s's}^\zeta({\bf k},{\bf q}).
\end{align}
Here, the conservation law for momentum flowing into and out of the
interaction has been applied, $\kappa$ is the dielectric constant
for a ML-TMDs material, ${\bf q}=(q_x,q_y)$ is the change of the
electron wavevector during an e-e scattering event,  $V_q=2\pi
e^2/(\kappa q)$ is the 2D Fourier transformation of the Coulomb
potential induced by e-e interaction, and
\begin{align}
F_{ss'}^\zeta ({\bf k},{\bf q})=&[{\cal A}_{s'}^\zeta ({\bf k+q})
{\cal A}_{s}^\zeta ({\bf k})]^2\sum_{i=1}^4 c_{is}^{\zeta *} ({\bf k})
c_{is'}^{\zeta}({\bf k+ {\bf
q}})\nonumber\\
&\times\sum_{j=1}^4 c_{js'}^{\zeta*} ({\bf k}+
{\bf q}) c_{js}^{\zeta}({\bf k}),
\end{align}
is the form factor for many-body interaction, where
${\cal A}_{s}^\zeta ({\bf k})={\cal A}_{c,s}^\zeta ({\bf k})$ and the
hybridization of the four electron wave functions, given by
Eq. \eqref{eigenfunctions}, at a state $|{\bf k};\zeta,s\rangle$ has
been included.

With the electrostatic potential induced by bare e-e interaction, the
dynamical dielectric function for electrons in spin split
conduction band in valley $\zeta$ can
be calculated by the RPA approach, which is given by a $2\times 2$ matrix as
\begin{equation}
\varepsilon_\zeta (\Omega,{\bf q})= \left[\begin{array}{cc}
 1-A_{--}\ \ & A_{-+} \\ A_{+-} \ \ & 1-A_{++}\\
 \end{array}\right],\end{equation}
where $A_{ss'}=A_{ss'}^\zeta (\Omega,{\bf q})=V_q \sum_{\bf k}
F_{ss'}^\zeta ({\bf k}, {\bf q}) \Pi_{ss'}^\zeta (\Omega; {\bf
k},{\bf q})$ is for scattering of an electron in a split state $s$
to a split state $s'$, $\Omega$ is the excitation frequency, and
\begin{equation}
\Pi_{s's}^\zeta (\Omega; {\bf k},{\bf q})={f[E_{s'}^\zeta({\bf
k}+{\bf q})]-f[E_{s}^\zeta({\bf k})]\over E_{s'}^\zeta({\bf k}+{\bf
q})-E_{s}^\zeta({\bf k})+\hbar\Omega+i\delta},
\end{equation}
is the corresponding density-density (d-d) correlation function (or
pair bubble) with $f(x)=[e^{(x-E_F)/k_BT}+1]^{-1}$ being the
Fermi-Dirac function and $E_F$ the Fermi energy or chemical
potential.

In a static case ($\Omega\to 0$) and long-wavelength limit ($q\to 0$),
the real part of the d-d correlation function becomes
\begin{equation}
D_{ss}^\zeta ({\bf k})
=\lim_{q\to 0}{\rm Re}\Pi_{ss}^\zeta (0; {\bf k},{\bf q})
\simeq {\partial f(x) \over \partial x}\Big|_{x=E_{s}^\zeta({\bf
k})},
\end{equation}
 for intra-subband ($s'=s$) transition, and
\begin{equation}
D_{s's}^\zeta ({\bf k})
=\lim_{q\to 0}{\rm Re}\Pi_{s's}^\zeta (0; {\bf k},{\bf q})
\simeq {f[E_{s'}^\zeta({\bf k})]-f[E_{s}^\zeta({\bf k})]\over
E_{s'}^\zeta({\bf k})-E_{s}^\zeta({\bf k})},
\end{equation}
for inter-subband ($s\neq s'$) transition, respectively.

By definition, the effective e-e interaction potential in the
presence of electronic screening can be calculated through a matrix:
\begin{equation}
[{\cal V}_{ss'}^\zeta({\bf k},{\bf q})]=[{V}_{ss'}^\zeta({\bf
k},{\bf q})][{\rm Re}\ \varepsilon_\zeta (0,{\bf q})]^{-1},
\end{equation}
which reads
\begin{widetext}
\begin{align}\label{effectivee-e}
\setlength{\arraycolsep}{0.2pt}
[{\cal V}_{ss'}^\zeta({\bf k},{\bf q})]&=\left[\begin{array}{cc}
{\cal V}_{--}\ \ & {\cal V}_{-+} \\ {\cal V}_{+-} \ \ & {\cal V}_{++} \\
 \end{array}\right]
=\left[\begin{array}{cc}{V}_{--}\ \ & {V}_{-+} \\ {V}_{+-}
\ \ & {V}_{++} \\
 \end{array}\right]
\left[\begin{array}{cc}
 1-A_{--}^{(1)}\ \ & A_{-+}^{(1)} \\ A_{+-}^{(1)} \ \ & 1-A_{++}^{(1)} \\
 \end{array}\right]^{-1}\nonumber\\
&={2\pi e^2\over \kappa {\cal Q}^2}
\left[\begin{array}{cc}
 F_{--}^{\zeta}(q+K_{++}^{\zeta})+F_{-+}^{\zeta}(K_{+-}^\zeta)\ \ \ \ & F_{-+}^{\zeta} (q+K_{--}^{\zeta})+
 F_{--}^{\zeta}(K_{-+}^\zeta)\\ F_{+-}^{\zeta} (q+K_{++}^\zeta)+F_{++}^{\zeta}(K_{+-}^\zeta) \ \
 \ \
 & F_{++}^{\zeta}(q+K_{--}^\zeta)+F_{+-}^{\zeta}(K_{-+}^\zeta) \\
 \end{array}\right],
\end{align}
\end{widetext}
where $A_{ss'}^{(1)}={\rm Re}A_{ss'}=V_q \sum_{\bf k} F_{ss'}^\zeta
({\bf k}, {\bf q}){\rm Re}\Pi_{s's}^\zeta (0;{\bf k}, {\bf q})$ and
${\cal Q}^2=(q+K_{--})(q+K_{++})-K_{-+}K_{+-}$. The inverse
static screening length is defined as
\begin{align}\label{screeninglength}
K_{ss'}^\zeta &=K_{ss'}^\zeta({\bf q})=-q V_q \sum_{\bf k} F_{ss'}^\zeta ({\bf k},
{\bf q}) {\rm Re}\Pi_{ss'}^\zeta (0,{\bf k},{\bf q})\nonumber\\
&=-{2\pi e^2\over \kappa} \sum_{\bf k} F_{ss'}^\zeta ({\bf k}, {\bf q}) {\rm
Re}\Pi_{ss'}^\zeta (0,{\bf k},{\bf q}),
\end{align}
which implies that different electronic transition channels correspond
to different screening lengths. Eq. \eqref{screeninglength} reflects a fact
that in the presence of electronic screening the effective electronic
transition from $s$ to $s'$ spin states should, in principle, be
affected by other transition events. In the long-wavelength limit
($q\to 0$) we have $F_{ss}^\zeta ({\bf k}, {\bf q})\to 1$ for $s=s'$
because of the normalization of the wave function, and
$F_{ss'}^\zeta ({\bf k}, {\bf q})\to 0 $ and $K_{ss'}^\zeta({\bf
q})\to 0$ for $s\neq s'$ because of the orthogonality of the electron wave
function. In such a case, Eq. \eqref{effectivee-e} becomes
\begin{equation}\label{11}
\setlength{\arraycolsep}{0.1pt}
\left[\begin{array}{cc}
{\cal V}_{--}\ \ & {\cal V}_{-+} \\ {\cal V}_{+-} \ \ & {\cal V}_{++} \\
 \end{array}\right]
 ={2\pi e^2\over \kappa}\left[\begin{array}{cc}
 (q+K_{--}^\zeta)^{-1}\ \ \ \ & 0 \\ 0 \ \
 \ \
 & (q+K_{++}^\zeta)^{-1} \\
 \end{array}\right],
\end{equation}
where
\begin{equation}\label{slength}
K_{ss}^\zeta=-(2\pi e^2/\kappa) \sum_{\bf k} [{\partial f(x) /\partial
x}]|_{x=E_{s}^\zeta({\bf k})},
\end{equation}
is the inverse screening length which is independent of ${\bf k}$
and ${\bf q}$. This result indicates that in the long-wavelength limit
the effective e-e interaction and the electronic screening can only
be caused via intra-subband electronic transitions. The inter-subband
transitions, corresponding to spin-flip transitions, can only be
achieved with the change of electron momentum during the scattering
events.

\subsection{Electronic transport coefficients}
In this study, we employ a simple Boltzmann equation (BE) approach
to calculate the transport coefficient for a ML-TMDs in the presence
of proximity induced interactions. In the present study, we neglect
the electronic transitions between different valleys, because these
transition channels require a big change of electron momentum that
is less possible in the transport experiment under the action of a
relatively weak driving dc electric field. For an $n$-type ML-TMDs
with a spin splitting conduction band, a two-band model is required to describe
the electronic properties in splitting bands. The time-independent
semi-classical Boltzmann equation can be written as
\begin{equation}\label{Boltzmannequation}
-\frac{{e}}{{\hbar}}{\bf F}\cdot\nabla_{\bf k} f^{\zeta}_{s}(\bf k)
=\sum_{\bf k',s'}[F^{\zeta}_{s',s}(\bf k',\bf k)-F^{\zeta}_{s',s}(\bf k,\bf k')],
\end{equation}
where $\bf F$ is a force acting on the electron, $f_{s}^{\zeta}(\bf k)$
is the momentum distribution function (MDF) for an electron in a state
$\mid {\bf k};\zeta,s\rangle$, $F_{ss'}^{\zeta}({\bf k},{\bf k}')
=f_{s}^{\zeta}({\bf k})W^{\zeta}_{ss'}({\bf k},{\bf k}')$, and
$W^{\zeta}_{ss'}(\bf k,\bf k')$ is the electronic transition rate for
scattering of an electron from a state $\mid {\bf k};\zeta,s\rangle$ to a
state $\mid {\bf k}';\zeta,s'\rangle$ in the conduction band due to the
presence of electronic scattering centers such as impurities and phonons.
When an electric field is applied along the $x$-direction of the ML-TMDs,
${\bf F}=F(1,0,0)$ is the strength of the dc electric field and we obtain
\begin{equation}
{\bf F}\cdot\nabla_{\bf k}f_s^{\zeta}({\bf k})=
F\frac{\partial f_{s}^{\zeta}({\bf k})}{\partial k_{x}}.
\end{equation}
For the first moment, the momentum-balance equation can be derived
by multiplying $\sum_{s,{\bf k}}(k_{x},k_{y})$ to both sides of
the Boltzmann equation given by Eq. \eqref{Boltzmannequation}, which reads
\begin{equation}\label{momentum-balance}
\frac{{en_{e}^{\zeta}}}{{\hbar}}(F,0)=\sum_{s',s,{\bf k'},{\bf k}}
(k'_{x}-k_{x},k'_{y}-k_{y})F_{ss'}^{\zeta}({\bf k},{\bf k'}),
\end{equation}
where $n_{e}^{\zeta}=\sum_{s,{\bf k}}f_{s}^{\zeta}({\bf k})$ is
the electron density in valley $\zeta$. It should be noted that
the main effect of driving electric field $F$ is to cause the drift
velocities of the electrons in different bands
$\bm v_{s}^{\zeta}=(v_{sx}^{\zeta},v_{sy}^{\zeta})$. As a result,
the electron wave vector in the MDF is shifted by
${\bf k}\rightarrow{\bf k}_{s}^{\zeta}={\bf k}-m_{s}^{\zeta}{\bf v}_s^{\zeta}/\hbar$,
with $m_{s}^{\zeta}$ being the transport effective mass for an
electron in the $(s,\zeta)$ band. As we know, the electrons
in a solid can be accelerated by a driving electric field.
Thus, the electrons in the bottom of the conduction band would move to
higher energy states with nonzero $k$. The electron effective mass under such a condition would
usually differ from the band mass in a parabolic low-
energy regime. This effective electron mass is often called the transport
effective mass. We note here that in general,
the transport effective mass for an electron differs from the band
effective mass $m^{\ast}$ obtained from taking
$1/m^{\ast}=(1/\hbar^2)d^{2}E/d^{2}k$ in an electronic system.
$m^{\ast}=m_s^{\zeta}$ normally holds for the case of
$E(k)=\hbar^{2}k^{2}/(2m^{\ast})$. For the case of a relatively
weak driving electric field $F$, the drift velocity of electron
is relatively small so that
\begin{equation}
f_s^{\zeta}({\bf k}_s^{\zeta})\simeq f_s^{\zeta}({\bf k})
-\frac{{ m_{s}^{\zeta}}}{{\hbar}}\bigg[v_{sx}^{\zeta}
\frac{{\partial f_{s}^{\zeta}({\bf k})}}{{\partial k_{x}}}
+v_{sy}^{\zeta}\frac{{\partial f_{s}^{\zeta}({\bf k})}}{{\partial k_{y}}}\bigg].
\end{equation}
Thus, we obtain from Eq. \eqref{momentum-balance} that
\begin{equation}\label{16}
(F,0)=-\sum_{s',s}\frac{{m_{s}^{\zeta}}}{e}(v_{sx}^{\zeta}
S_{ss'}^{\zeta x}+v_{sy}^{\zeta}T_{ss'}^{\zeta y},v_{sx}^{\zeta}
T_{ss'}^{\zeta x}+v_{sy}^{\zeta}S_{ss'}^{\zeta y}),
\end{equation}
where
\begin{align}\label{17}
[S_{ss'}^{\zeta \alpha},T_{ss'}^{\zeta \alpha}]
=&\frac{1}{n_{e}^{\zeta}}\sum_{{\bf k}',{\bf k}}(k'_{\alpha}-k_{\alpha})
W_{ss'}^{\zeta}({\bf k},{\bf k}')\nonumber\\
&\times\bigg[\frac{{\partial f_{s}^{\zeta}({\bf k})}}{{\partial k_{\alpha}}},
\frac{{\partial f_{s}^{\zeta}({\bf k})}}{{\partial k_{{\alpha}'}}}\bigg],
\end{align}
with $\alpha=x$ or $y$ and $\alpha'\neq\alpha$, presents
the scattering probability for an electron moving along different
directions in different spin and valley subbands .
Assuming that the
electron MDF can be described by a statistical energy distribution
function (EDF) such as the Fermi-Dirac function, we have
$f_{s}^{\zeta}({\bf k})\simeq f[E_{s}^{\zeta}({\bf k})]$ with
$f(x)=[e^{(x-E_{F})/k_{B}T}+1]^{-1}$. In an $n$-type ML-TMDs at the steady
state, the single Fermi level is across the system. Thus, we obtain
\begin{align}
[S_{ss'}^{\zeta\alpha},T_{ss'}^{\zeta\alpha}]
=&\frac{ 1}{ n_{e}^{\zeta}}\sum_{{\bf k}',{\bf k}}(k_{\alpha}'-k_{\alpha})
W_{ss'}^{\zeta}({\bf k},{\bf k}')\nonumber\\
&\times\bigg[\frac{{\partial E_{s}^{\zeta}({\bf k})}}{{\partial k_{\alpha}}},
\frac{{\partial E_{s}^{\zeta}({\bf k})}}{{\partial k_{{\alpha}'}}}\bigg]
\frac{{df(x)}}{{dx}}\big|_{x=E_{s}^{\zeta}({\bf k})}.
\end{align}
From Eq. \eqref{diagonalized}, we have
\begin{equation}
\frac{{\partial E_{s}^{\zeta}({\bf k})}}{\partial k_{\alpha}}
=A^{2}k_{\alpha}G_{s}^{\zeta}({\bf k}),
\end{equation}
with
\begin{equation}
G_{s}^{\zeta}({\bf k})=\frac{{4[E_{s}^{\zeta}({\bf k})]^{2}
-4A^{2}k^{2}-\Delta^{2}-d_{\zeta}^{c}d_{\zeta}^{v}}}{ {4[E_{s}^{\zeta}({\bf k})]^{3}
-2A_{2}E_{s}^{\zeta}({\bf k})+A_{1}} },
\end{equation}
and
\begin{align}\label{rate}
[S_{ss'}^{\zeta\alpha},T_{ss'}^{\zeta\alpha}]
=&\frac{ {A^{2}}}{{n_{e}^{\zeta}}}\sum_{{\bf k}',{\bf k}}
(k_{\alpha}'-k_{\alpha})[k_{\alpha},k_{{\alpha}'}]G_{s}^{\zeta}({\bf k}) \nonumber\\
&\times W_{ss'}^{\zeta}({\bf k},{\bf k}')\frac{{df(x)}}{{dx}}\big|_{x=E_{s}^{\zeta}({\bf k})}.
\end{align}
By definition, the current density for electrons in band $(s,\zeta)$ is $j_{s\alpha}^{\zeta}=-e^{2}n_{s}^{\zeta}v_{s\alpha}^{\zeta}$ along the
$\alpha$ direction, with $n_{s}^{\zeta}$ being the electron density at
$(\zeta,s)$ state. Using the Onsager relation, Eq. \eqref{rate} gives
\begin{align}
\left[\begin{array}{cc}\rho_{xx}^{\zeta}\ \ & \rho_{xy}^{\zeta}\\ \rho_{yx}^{\zeta}\ \ &\rho_{yy}^{\zeta}\end{array}\right]&=\sum_{s}\left[\begin{array}{cc}\rho_{xx}^{\zeta s}\ \ & \rho_{xy}^{\zeta s}\\ \rho_{yx}^{\zeta s}\ \ &\rho_{yy}^{\zeta s}\end{array}\right]\nonumber\\
&=\frac{{m_{s}^{\zeta}}}{{e^{2}n_{s}^{\zeta}}}\sum_{s,s'}\left[\begin{array}{cc}S_{ss'}^{\zeta x}\ \ & T_{ss'}^{\zeta x}\\T_{ss'}^{\zeta y}\ \ & S_{ss'}^{\zeta y}\end{array}\right],
\end{align}
where $\rho_{\alpha\alpha'}^{\zeta s}$ is the longitudinal $(\alpha=\alpha')$
or transverse $(\alpha\neq\alpha')$ resistivity in the band $(\zeta,s)$. Noting that $\rho_{\alpha\alpha'}^{\zeta s}=m_{s}^{\zeta}\lambda_{\alpha\alpha'}^{\zeta s}/(e^{2}n_{s}^{\zeta})$ with $\lambda_{\alpha\alpha'}^{\zeta s}$ being the electronic scattering rate in
the band $(\zeta,s)$ along different directions, we obtain
\begin{equation}\label{scatteringrate}
\left[\begin{array}{cc}\lambda_{xx}^{\zeta s}\ \ & \lambda_{xy}^{\zeta s}\\ \lambda_{yx}^{\zeta s}\ \ &\lambda_{yy}^{\zeta s}\end{array}\right]=\sum_{s'}\left[\begin{array}{cc}S_{ss'}^{\zeta x}\ \ & T_{ss'}^{\zeta x}\\ T_{ss'}^{\zeta y}\ \ &S_{ss'}^{\zeta y}\end{array}\right].
\end{equation}
The electron mobility is defined as $\mu_{\alpha\alpha'}^{\zeta s}
=e\tau_{\alpha\alpha'}^{\zeta s}/m_{s}^{\zeta}$, with
$\tau_{\alpha\alpha'}^{\zeta s}=1/\lambda_{\alpha\alpha'}^{\zeta s}$
being the momentum relaxation time or lifetime for an electron in
band $(\zeta,s)$ along different directions. Finally, the average
longitudinal and transverse or Hall mobilities are given by
\begin{align}\label{mobilityxx}
\mu_{xx}=\frac{n_{+}^{+}\mu_{xx}^{++}
+n_{+}^{-}\mu_{xx}^{-+}+n_{-}^{+}\mu_{xx}^{+-}+n_{-}^{-}\mu_{xx}^{--}}{n_{e}},
\end{align}
and
\begin{align}\label{mobilityxy}
\mu_{xy}=\frac{n_{+}^{-}\mu_{xy}^{-+}+n_{-}^{-}\mu_{xy}^{--}-n_{+}^{+}\mu_{xy}^{++}
-n_{-}^{+}\mu_{xy}^{+-}}{n_{e}},
\end{align}
respectively, where $n_e=\sum_{\zeta,s} n_s^\zeta$ is the total electron density
in the system. For average $\mu_{xy}$ the electric currents in different
directions with different valley indexes are considered. Here we take the standard definition of the elements in the mobility tensor as $\mu_{\alpha\alpha'}=v_{\alpha'}/F_\alpha$ with the electron drift velocity (or current) measured along the $\alpha'$-direction and the driving electric field strength (or applied voltage) along the $\alpha$-direction.

\subsection{Electron-impurity scattering}
At relatively low temperatures, the electron-impurity (e-i) scattering
is the principal channel for relaxation of electrons in an electronic
system in the presence of a driving electric field. For the case in which
the e-i scattering is achieved through the Coulomb potential induced by
charged impurities that are three-dimensional-like, the e-i interaction
Hamiltonian is given as:
\begin{equation}
H_{e-i}=e^{2}/(\kappa_{i}|\bm{R}-\bm{R_{i}}|),
\end{equation}
where ${\bf R}=({\bf r},0)=(x,y,0)$ is the coordinate of an electron in ML-TMDs,
the impurity is located at ${\bf R}_{i}=({\bf r}_{i},z_{i})=(x_{i},y_{i},z_{i})$,
and $\kappa_{i}$ is the static dielectric constant of the medium that
contains the impurities. After assuming that the system can be separated
into the electrons of interest $|{\bf k};\lambda\rangle$ and the rest of the impurities
$|I\rangle$, namely $|{\bf k};\lambda,I\rangle=|{\bf k};\lambda\rangle|I\rangle$, the e-i interaction
matrix element is obtained, in the absence of e-e screening, as \cite{Dong08}
\begin{align}\label{25}
U({\bf q},{\bf R}_{i})=&\langle{\bf k}';\lambda',I|H_{e-i}|{\bf k};\lambda,I\rangle\nonumber\\
=&\frac{{2\pi e^{2}}}{{\kappa_{i}q}}\sqrt{n_{i}(z_{i})}e^{i{\bf q}
\cdot{\bf r}_{i}}e^{-q|z_{i}|}H_{ss'}^{\zeta}({\bf k},{\bf k}')
\delta_{{\bf k}',{\bf k}+{\bf q}},
\end{align}
where $\langle I|I\rangle=[n_{i}(z_{i})]^{1/2}$, with $n_{i}(z_{i})$ being the impurity
distribution along the $z$ direction, ${\bf q}=(q_{x},q_{y})$ is the change
of the electron wave vector during an e-i scattering event, and
$H_{ss'}^{\zeta}({\bf k},{\bf k}')=\langle{\bf k}';\lambda'|{\bf k};\lambda\rangle
={A}_{s'}^{\zeta}({\bf k}'){A}_{s}^{\zeta}({\bf k})
\Sigma_{j=1}^{4}c_{js'}^{\zeta\ast}({\bf k}')c_{js}^{\zeta}({\bf k})$ is
the form factor for e-i scattering. Here we have assumed that the impurities
are distributed uniformly along the $x$-$y$ plane. Using Fermi's Golden Rule,
the electronic transition rate for scattering of an electron from a state
$|{\bf k};\zeta, s\rangle$ to a state $|{\bf k}';\zeta,s'\rangle$ due to e-i interaction
is obtained, in the presence of e-e screening, as \cite{Dong08}
\begin{align}\label{transitionrate}
W_{ss'}^{\zeta}({\bf k},{\bf k}')=&\frac{{2\pi}}{{\hbar}}N_{i}(q)
|U_{ss'}^{\zeta}(q)|^{2}|H_{ss'}^{\zeta}({\bf k},{\bf k}')|^{2}\nonumber\\
&\times\delta_{{\bf k}',{\bf k}+{\bf q}}\delta[E_{s}^{\zeta}({\bf k})-E_{s'}^{\zeta}({\bf k}')],
\end{align}
where $$U_{ss'}^{\zeta}(q)={2\pi e^{2}\over \kappa_{i}(q+K_{ss'}^{\zeta})},$$ is the
screened e-i interaction potential and $N_i(q)=\int dz_{i}n_{i}(z_{i})e^{-2q|z_{i}|}$.
When a ML-TMDs sheet is placed on a
dielectric or magnetic substrate, the background impurities in the  ML-TMDs
layer and the impurities in the substrate can contribute to the e-i
interaction. Normally, the concentrations of these impurities are very hard to
determine experimentally. To
reduce the fitting parameters for the theoretical study, here we assume that the
impurities are effectively located at the interface between the ML-TMDs and the
substrate with an effective concentration $N_{i}$, i.e., $n_{i}=N_{i}\delta(z)$.
Thus, $N_i(q)=N_i$ is the areal concentration of the impurities.

Now we consider the case of low temperatures with $T\rightarrow0$.
In such a case, we have $df(x)/dx=-\delta(E_{F}-x)$ and $f(x)=\Theta(x)$,
with $\Theta(x)$ being the unit step-function. The condition of electron
number conservation leads to $n_{s}^{\zeta}=\sum_{{\bf k}}f[E_{s}^{\zeta}({\bf k})]$
so that the Fermi wavevector at the $(\zeta,s)$ band is
$k_{F}^{\zeta s}=[4\pi n_{s}^{\zeta}]^{1/2}$, which is the solution of $k$ from
$E_{F}-E_{s}^{\zeta}({\bf k})=0$. Introducing the electronic transition rate
induced by e-i interaction, given by Eq. \eqref{transitionrate}, into
Eq. \eqref{scatteringrate}, we obtained
\begin{align}\label{27}
\left[\begin{array}{cc}
S_{ss'}^{\zeta x}\ \  T_{ss'}^{\zeta x}\\ T_{ss'}^{\zeta y}\ \ S_{ss'}^{\zeta y}\end{array}\right]=&\frac{{N_{i}n_{s}^{\zeta}}}{{2\pi^{2}\hbar n_{e}^{\zeta}A^{2}}}\frac{{{\rm sign}[G(n_{s}^{\zeta})]}}{{G(n_{s'}^{\zeta})}}\int_{0}^{2\pi}d\theta\int_{0}^{2\pi}
d\phi\nonumber\\
&\times|U_{ss'}^{\zeta}(q)|^{2}|H_{ss'}^{\zeta}({\bf k}',{\bf k})|^{2}[R(\phi,\theta)],
\end{align}
where sign$(x)$ is the sign function,
\begin{equation}
G(n_{s}^{\zeta})=\frac{{4E_{F}^{2}-16\pi A^{2}n_{s}^{\zeta}-\Delta^{2}-d_{\zeta}^{c}d_{\zeta}^{v}}}{{4E_{F}^{3}-2A_{2}E_{F}+A_{1}}},
\end{equation}
and
\begin{align}\label{29}
\setlength{\arraycolsep}{0.1pt}
[R(\phi,\theta)]=
\left[\begin{array}{cc}
p_1{\rm cos}\phi\ \ & p_1{\rm sin}\phi \\
p_2{\rm cos}\phi\ \
& p_2{\rm sin}\phi\end{array}\right],
\end{align}
where $p_1=\gamma_{ss'}^{\zeta}{\rm cos}(\theta+\phi)-{\rm cos}\phi$,
$p_2=\gamma_{ss'}^{\zeta}{\rm sin}(\theta+\phi)-{\rm sin}\phi$,
with $\gamma_{ss'}^{\zeta}=(n_{s'}^{\zeta}/n_{s}^{\zeta})^{1/2}$.
Here, ${\bf k'}=(4\pi n_{s'}^{\zeta})^{1/2}[{\rm cos}(\theta+\phi),
{\rm sin}(\theta+\phi)]$, $ {\bf k}=(4\pi n_{s}^{\zeta})^{1/2}[{\rm cos}\phi,{\rm sin}\phi]$,
$q=2\sqrt{\pi}[n_{s'}^{\zeta}+n_{s}^{\zeta}-2(n_{s'}^{\zeta}n_{s}^{\zeta})^{1/2}{\rm cos}\theta]^{1/2}$
with $\theta$ being the angle between ${\bf k'}$ and $\bf k$.

\section{Results and discussions}
\label{sec:results and discussions}

In this study, we take $n$-type ML-MoS$_2$ as an example to look
into the influence of the proximity effect on electronic and transport
properties of the 2D TMDs system. The material and theoretical parameters
for ML-MoS$_2$ are taken as \cite{Xiao12,Zhu11,Lu13,Li12}: $A=at=3.5123$
eV{\AA} where $a=3.193$ {\AA} and $t=1.1$ eV, $\Delta=1.66$ eV,
$\lambda_{c}=-1.5$ meV, and $\lambda_{v}=75$ meV. We note that at present,
the experimental and theoretical results of the transport effective electron
masses for an electron at the states $(\zeta,s)$, $m_s^\zeta$, are
unavailable. Because ML-MoS$_2$ has a roughly parabolic band structure around the $K$ and $K'$ points \cite{Zhao20},
here we assume that they are not different far from the
average density-of-state effective electron mass, namely we assume
$m_s^\zeta \simeq m_{e}=0.52 m_{0}$ for ML-MoS$_2$ \cite{Li14}, with
$m_0$ being the rest electron mass. We consider an air/ML-MoS$_{2}$/EuO
system where EuO is the magnetic substrate, which can result in the effect
of proximity-induced exchange interaction. The dielectric constants for
air, a bare ML MoS$_{2}$ sheet, and a bare EuO substrate are taken to be, respectively, $\kappa_{air}= 1$, $\kappa_{\mathrm{TMD}}=3.3$ \cite{Paul22} and
$\kappa_{Sub}=23.9$ \cite{Leroux-Hugon72}. Considering the mismatch of
the dielectric constants at the ML-MoS$_{2}$/EuO interface, we evaluate
the effective dielectric constants for ML-MoS$_{2}$ and the substrate
from the bare dielectric constants using the mirror image method \cite{Xu09}.
Thus, we have $\kappa = (\kappa_{air} + \kappa_{TMD} )/2 = 2.15$ for
ML-MoS$_{2}$ and $\kappa_{i} = (\kappa_{TMD} + \kappa_{Sub} )/2 = 13.6$
for the substrate. Because the strength of the Rashba parameter, $\lambda_{R}$,
and the effective Zeeman fields for conduction and valence bands, $B_{c}$ and
$B_{v}$, can be tuned experimentally, we take them as variable input parameters
in numerical calculations. It should be noted that the strengths of $B_{c}$ and $B_{v}$
are usually different with different effective Land\'{e} $g$-factors for
Bloch states \cite{Qi15}. Because our attention in this study is mainly
given to the conduction band, we take $B_v=5$ meV in all calculations. Since the impurity concentration
in a ML-MoS$_2$/substrate system, $n_i$, is normally unknown, we take it as a fitting parameter that
can be determined by, e.g., using the experimental data of the sample
mobility. Furthermore, for a given total electron density in the sample system,
$n_e$, we can determine the Fermi energy $E_F$ by using the condition of
electron number conservation,
\begin{equation}
 n_e=\sum_{s,\zeta,{\bf k}}f[E_s^\zeta ({\bf k})].
\end{equation}
With the obtained Fermi energy, the electron density at a spin state $s$
in valley $\zeta$ in the conduction band can then be calculated via
\begin{equation}
 n_s^\zeta=\sum_{{\bf k}}f[E_s^\zeta ({\bf k})].
\end{equation}

\subsection{Electronic band structure and Berry curvature}

\begin{figure}[t]
\centering
\includegraphics[width=8.6cm]{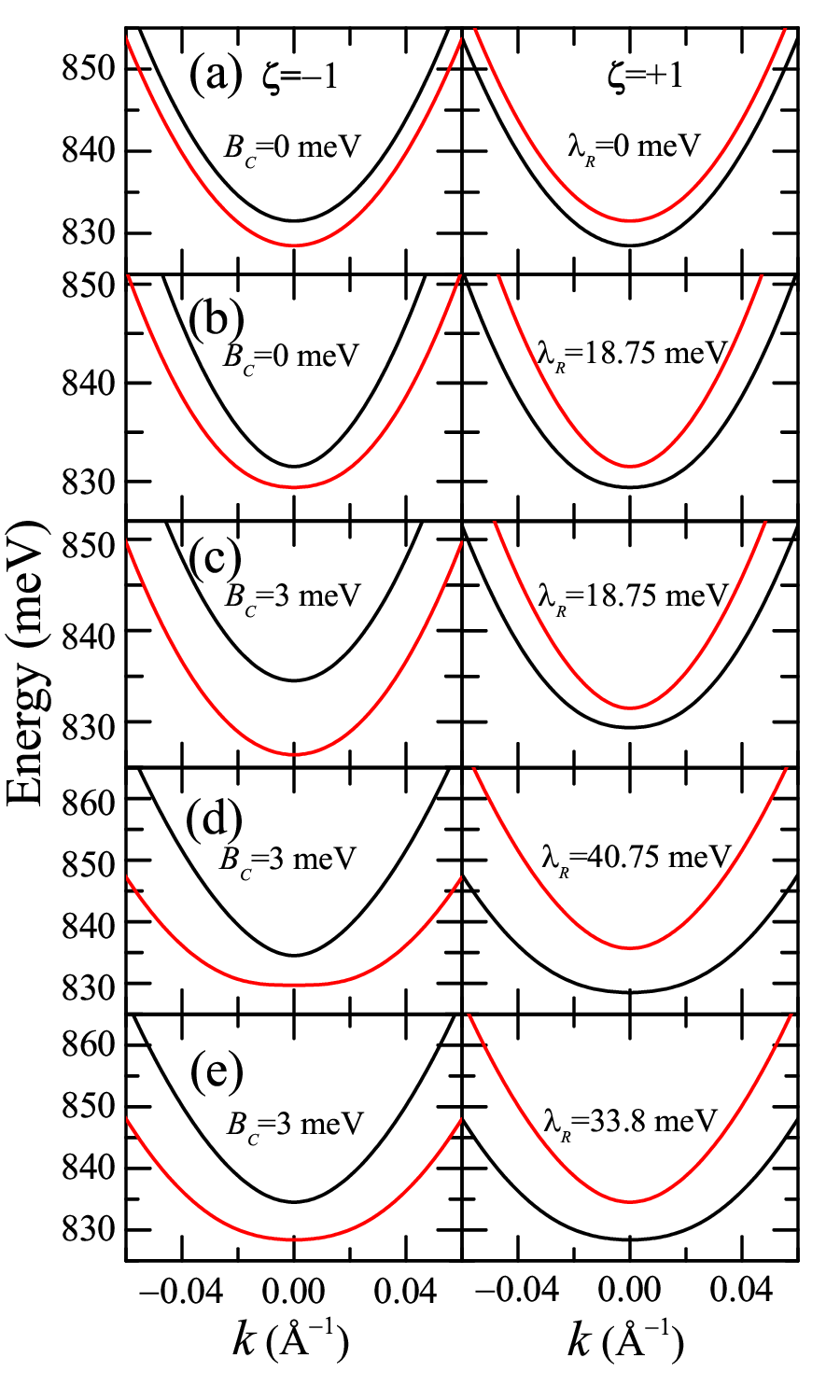}
\caption{The electronic band structure of ML-MoS$_{2}$ in the presence of proximity
induced interactions for conduction subbands in two valleys ($\zeta=\pm$1).
The spin up/down states are represented by red and black curves, respectively.
The results with different values of $\lambda_{R}$ and $B_{c}$ as indicated
are shown in panels (a)-(e).}
\label{fig1}
\end{figure}

Some of the basic features of the electronic band structure in ML-MoS$_2$
have been discussed in our previous research \cite{Zhao20}. For convenience
of understanding the influence of proximity-induced interactions on electronic
and transport properties of ML-MoS$_2$, to be presented and discussed later, here we present
some relevant results about how the Rashba parameter and the strength of EZF
would affect the electronic energy levels in the conduction band in ML-MoS$_2$.
In Fig. \ref{fig1} we show the energy levels in a spin-split ($s=\pm$) conduction
band in different valleys ($\zeta=\pm 1$) for different Rashba parameters
$\lambda_{R}$ and EZF factors $B_{c}$. We notice the following features. (i)
When $B_c=0$ the conduction bands in different valleys degenerate (see Figs. \ref{fig1}(a)
and (b)). When $\lambda_R=0$ the splitting of the conduction band is induced
by intrinsic SOC $\lambda_c$ (see Fig. \ref{fig1}(a)). (ii) When $B_c\neq 0$, the degeneracy
of the energy levels in different valleys is lifted (see Figs. \ref{fig1}(c) and (d)). (iii)
At a fixed $B_c$, the energy spacing between two split levels in a certain valley
does not increase monotonously with $\lambda_R$ and a different spin effect can
be observed in different valleys (see Figs. \ref{fig1}(c) and (d)). (iv) Interestingly,
at a fixed $B_c=3$ meV, the minimum of the conduction band can be seen in
the $\zeta=-1$ valley (see Fig. \ref{fig1}(c)) when $\lambda_R=18.75$ meV, whereas it can be seen in the $\zeta=+1$ valley (see Fig. \ref{fig1}(d)) when
$\lambda_R=40.75$ meV. And (v) For $\lambda_{R}=33.8$ meV in Fig. \ref{fig1}(e), the
electronic structure is roughly degenerate for both valleys.
These findings indicate that in the presence of proximity-induced
interactions, the electronic band structure in ML-MoS$_2$ depends strongly on
$\lambda_R$ and $B_c$.

\begin{figure}[t]
\centering
\includegraphics[width=8.6cm]{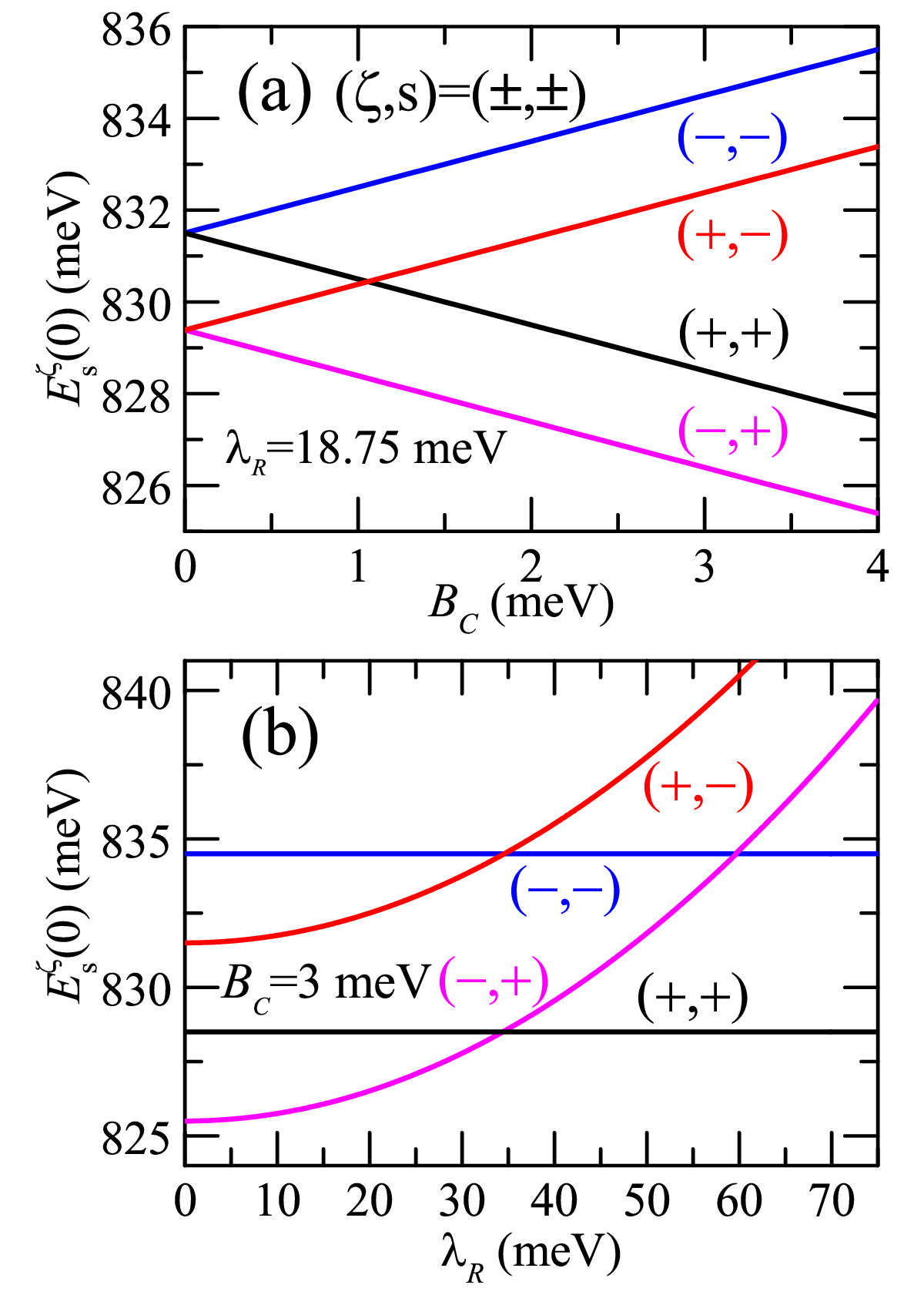}
\caption{The energies of the bottoms of four conduction subbands, $E_{s}^{\zeta}(0)$
with $k=0$ and $(\zeta,s)=(\pm,\pm)$, as a function of $B_{c}$ at a fixed $\lambda_{R}=18.75$
meV in (a) and as a function of $\lambda_{R}$ at a fixed $B_{c}=3$ meV in (b).}
\label{fig2}
\end{figure}

\begin{figure}[t]
\centering
\includegraphics[width=8.6cm]{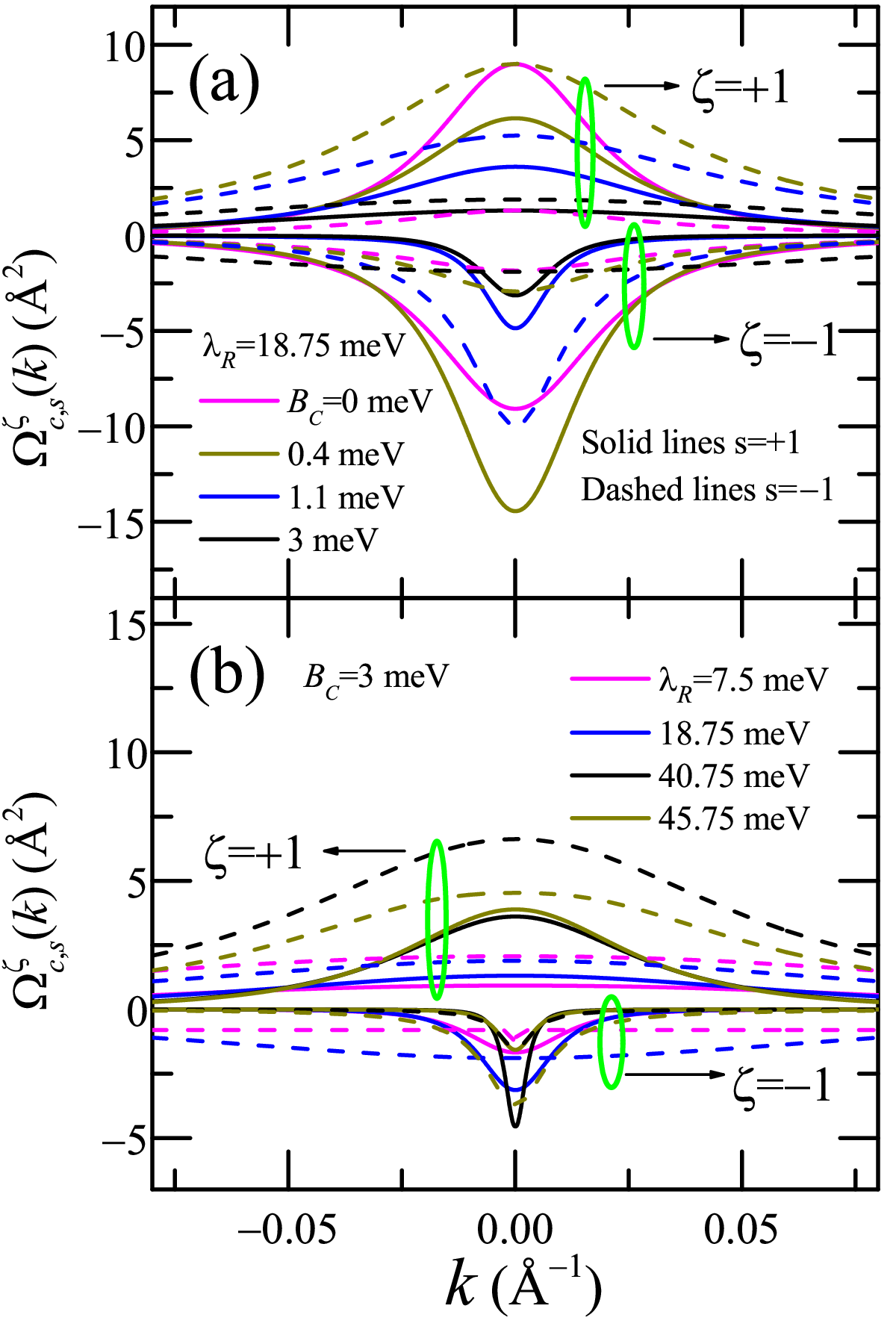}
\caption{The Berry curvatures as a function of wavevector $k$ for conduction bands
(a) at fixed $\lambda_{R}=18.75$ meV with different EZF $B_{c}$, and (b) at
fixed $B_{c}=3$ meV with different Rashba SOC $\lambda_{R}$ near $K$ and $K'$ points.
The solid and dashed lines refer to the results of spin up and spin down subbands
and the green circles are collected for different valleys.}
\label{fig3}
\end{figure}

In Fig. \ref{fig2}(a) we plot the bottoms of four conduction subbands,
$E_{s}^{\zeta}(0)$  with $k=0$ and $(\zeta,s)=(\pm,\pm)$, as a function of
$B_{c}$ at a fixed Rashba parameter $\lambda_{R}=18.75$ meV. We can see that
the effective Zeeman field $B_c$ can lead to different energy gaps between
spin-split conduction subbands in different valleys. In particular,
$E_{\pm}^{\pm}(0)$ for four conduction subbands varies linearly with increasing $B_{c}$.
With increasing $B_c$, the spin splitting of the conduction band at $K'$ point
$\zeta=-1$ always increases, whereas the spin splitting of the conduction subands
at $K$ point $\zeta=+1$ first decreases to zero and then increases. $B_{c}=1.1$ meV
is a peculiar point at which the energy levels of different spin subbands at
the $K$ point are flipped. These results indicate that the effective Zeeman field
factor $B_{c}$ can effectively tune the electronic band structure in the conduction
subbands in different valleys; in particular, the energy difference $E_{+}^{-}(0)-E_{+}^{+}(0)$
can be flipped via varying $B_c$.

In Fig. \ref{fig2}(b), $E_{s}^{\zeta}(0)$ for four conduction subbands are
shown as a function of $\lambda_R$ at a fixed $B_c=3$ meV. We find that
$E_{-}^{-}(0)$ and $E_{+}^{+}(0)$ in different valleys depend very weakly
on $\lambda_R$, whereas $E_{-}^{+}(0)$ and $E_{+}^{-}(0)$ in different
valleys increase rather rapidly with $\lambda_R$. This implies that the
RSOC or $\lambda_R$ affects mainly the spin-up levels in different valleys.
The presence of the proximity-induced exchange interaction
(i.e., $B_{c}\neq 0$) can lift the valley degeneracy and modify the spin
splitting. With increasing $\lambda_{R}$, the energy difference between
$E_{-}^{-}(0)-E_{+}^{-}(0)$ first decreases to zero and then increases,
and the energy difference between $E_{-}^{+}(0)-E_{+}^{+}(0)$ always increases.
There is also a peculiar point around $\lambda_{R}=33.8$ meV at which both
$E_{-}^{+}(0)-E_{-}^{-}(0)$ and $E_{+}^{-}(0)-E_{+}^{+}(0)$ approach  zero
and the lowest conduction subband is changed from $(\zeta,s)=(-,+)$ to
$(\zeta,s)=(+,-)$. This effect implies that the lowest conduction subband
can be changed from valley $K'$ to valley $K$ via varying the value of $\lambda_R$.
The results shown in Fig. \ref{fig2} suggest that the proximity-induced RSOC
and Zeeman effect can  modulate strongly the spin-split conduction band in
different valleys in ML-MoS$_2$.

As we know, in the presence of an applied electric field $\bf E$ along
ML-MoS$_2$ film, Berry curvature enters into the semiclassical
wavepacket dynamics, and the electrons would gains an anomalous
velocity \cite{Xiao10,Cao12}$(v_{sy}^{\zeta}\sim{\bf E}\times{\bf\Omega}^\zeta_{c,s}({\bf k}))$
perpendicular to the applied electric field. Thus, the Berry curvature
would play an important role in affecting the transverse or Hall currents.
In Fig. \ref{fig3}, we plot the Berry curvature ${\bf \Omega}^\zeta_{c,s}({\bf k})$
as a function of wavevector $\bf k$ at valleys $K$ and $K'$ with different
EZF and Rashba SOC. As we can see, the EZF $B_c$ and Rashba SOC $\lambda_R$
can significantly affect the values of Berry curvature. However, the values
of Berry curvature in valley $K$ and valley $K'$ are always positive and negative
with different EZF and Rashba parameters. This results implies that the directions
of the transverse currents contributed by different valleys remain unchanged. The
total transverse current is the summations of the contributions from two valleys
and the strength of the transverse current could be tuned by varying the EZF and
Rashba parameters.

\subsection{Electronic screening length}
\begin{figure}[t]
\centering
\includegraphics[width=8.6cm]{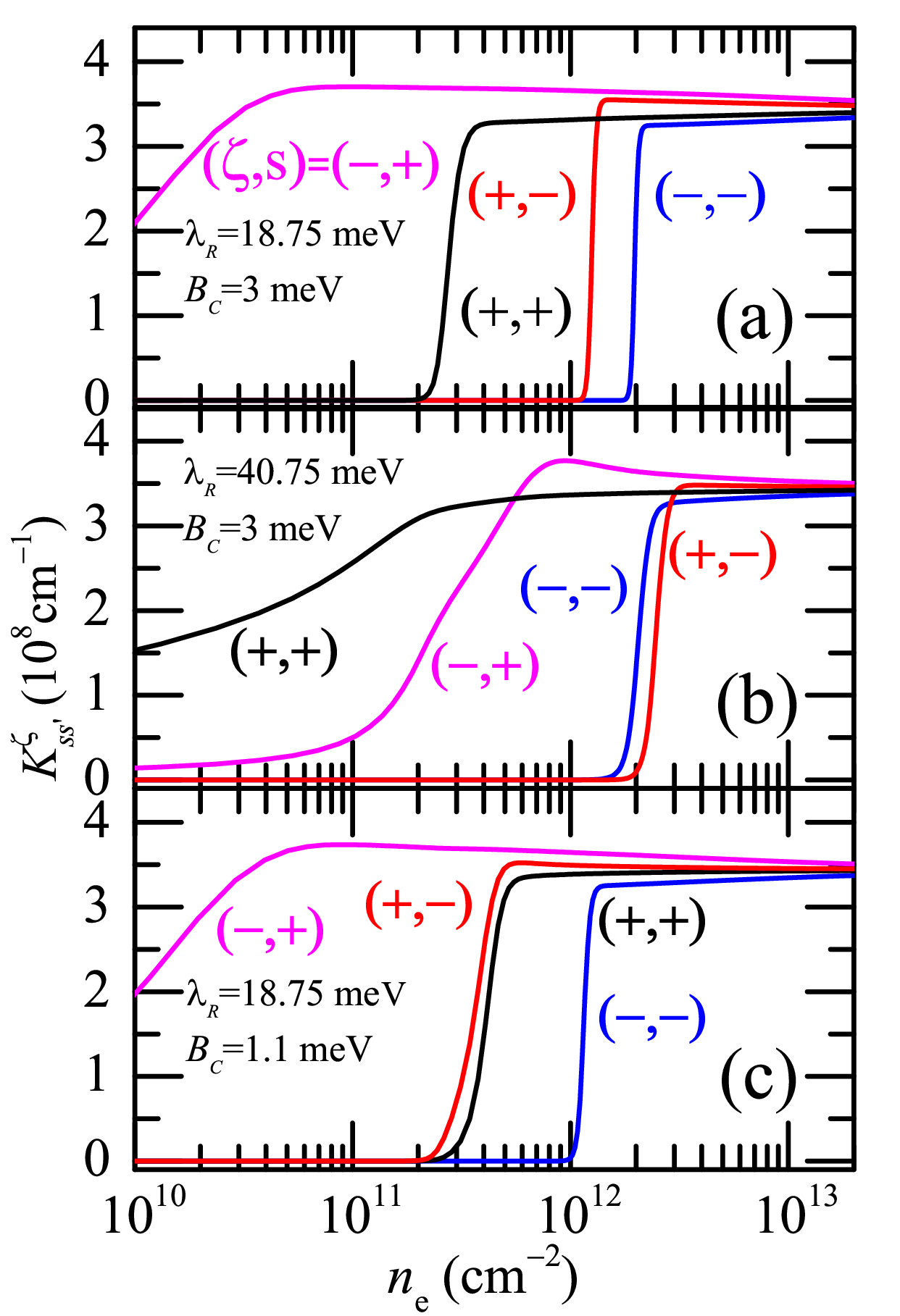}
\caption{The inverse screening lengthes $K_{ss'}^\zeta$ for intra-subband
transitions in four spin-split conduction subbands as a
function of electron density $n_{e}$ at $T\rightarrow0$ for
a fixed value of $B_v=5$ meV. Here,
$\lambda_{R}=18.75$ meV and $B_{c}=3$ meV in (a), $\lambda_{R}=40.75$
meV and $B_{c}=3$ meV in (b), and $\lambda_{R}=18.75$ meV and
$B_{c}=1.1$ meV in (c).}
\label{fig4}
\end{figure}

As shown in Eq. \eqref{screeninglength}, at $T\rightarrow0$ and $q\rightarrow0$
limits, the RPA inverse screening length $K_{ss'}^\zeta$ is attributed mainly to
intra-subband electronic transitions (i.e., $s'=s$). In this study, we only evaluate
the electronic screening induced by intra-subband e-e interaction by using Eq. \eqref{slength}.
In Fig. \ref{fig4}, we plot the inverse screening lengthes $K_{ss'}^\zeta$ for four
spin-split conduction subbands as a function of electron density $n_{e}$ at the
fixed valence EZF parameter $B_{v}=5$ meV and at the low-temperature limit
$T\rightarrow0$ K. Here, different values of $B_c$ and $\lambda_R$ are used to
examine the effect of $B_c$ and $\lambda_R$ on $K_{ss'}^\zeta$. We can see that
with increasing electron density, the effect of electronic screening first
increases and then depends relatively weakly on $n_e$, in agreement with the screening
effect found in, e.g., graphene \cite{Dong08}. $K_{ss'}^\zeta$ differs in spin-split
subbands in different valleys. Because $(\zeta,s)=(-,+)$, $(+,+)$ and $(-,+)$ are
the lowest electronic subbands (see Fig. \ref{fig2}) for corresponding parameters
indicated in Fig. \ref{fig4}, which are always occupied by electrons, $K_{++}^-$
in Fig. \ref{fig4}(a), $K_{++}^+$ in Fig. \ref{fig4}(b) and $K_{++}^-$ in
Fig. \ref{fig4}(c) are always non-zero values. The non-zero values of other
$K_{ss}^\zeta$ can only be obtained with increasing $n_e$ when the corresponding
higher subband becomes populated. At relatively large $n_e$ so that four conduction
subbands in both $K$- and $K'$-valleys are occupied by electrons, the inverse
screening lengths $K_{ss}^\zeta$ for four conduction subbands are approaching
 roughly the same value with increasing electron density. We note that because
the electronic energy spectrum for ML-TMDs is largely parabolic (see Fig. \ref{fig1}),
the dependence of $K_{ss}^\zeta$ upon $n_e$ for ML-TMDs is similar to those obtained
for semiconductor-based 2D electron gas (2DEG) systems \cite{Xu05}. For ML-MoS$_2$,
the inverse electronic screening length is in the order of $10^{8}$ cm$^{-1}$, which
is in line with the results obtained in semiconductor-based 2DEG systems \cite{Xu05}.

\subsection{Longitudinal and transverse mobilities}

\begin{figure}[t]
\includegraphics[width=8.6cm]{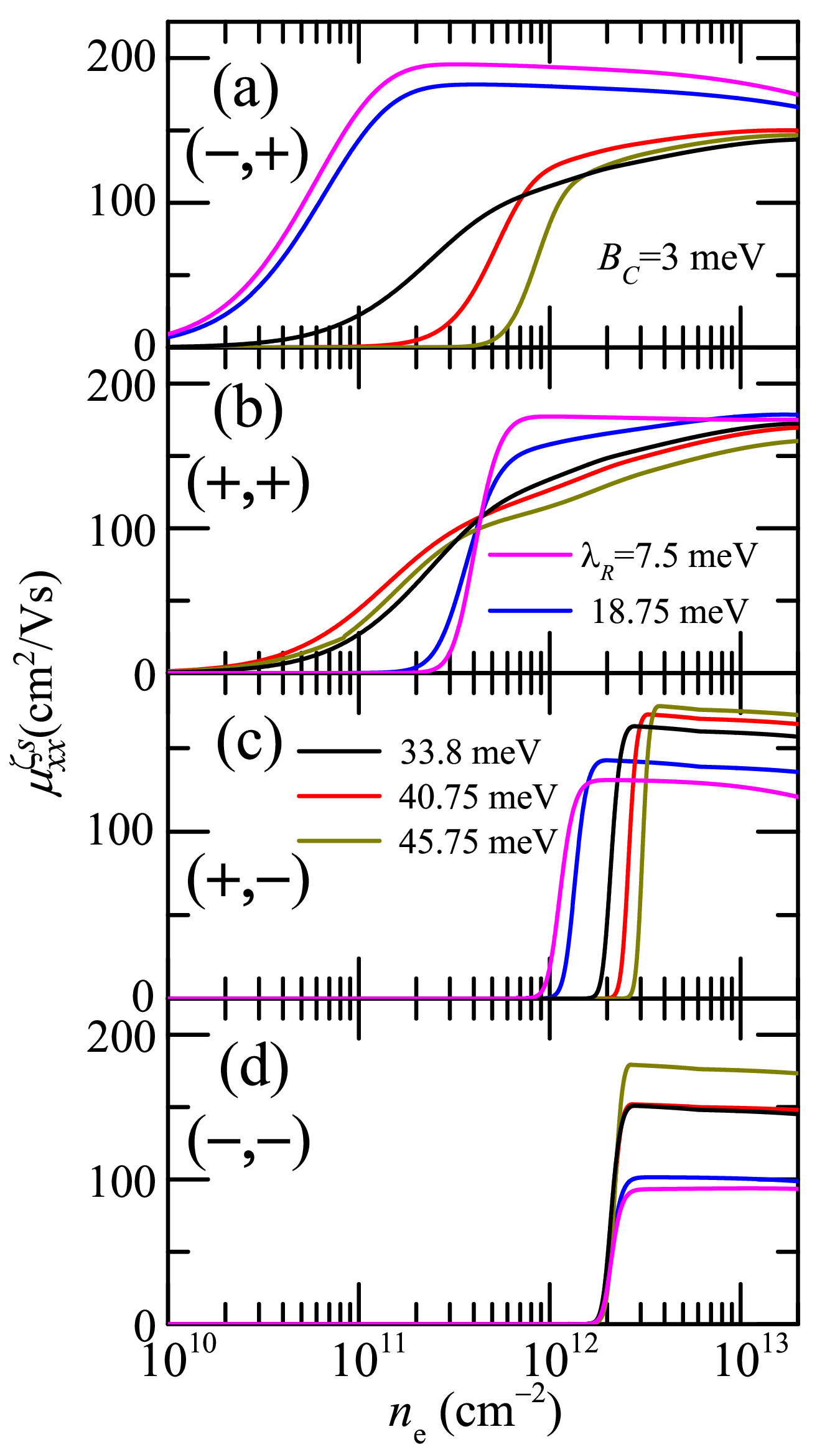}
\caption{The longitudinal mobility $\mu_{xx}^{\zeta s}$ in conduction
subband $(\zeta,s)=(\pm,\pm)$ in ML-MoS$_{2}$ as a function of electron
density at a fixed $B_c=3$ meV for different Rashba parameters $\lambda_{R}$
as indicated.}
\label{fig5}
\end{figure}

\begin{figure}[t]
\includegraphics[width=8.6cm]{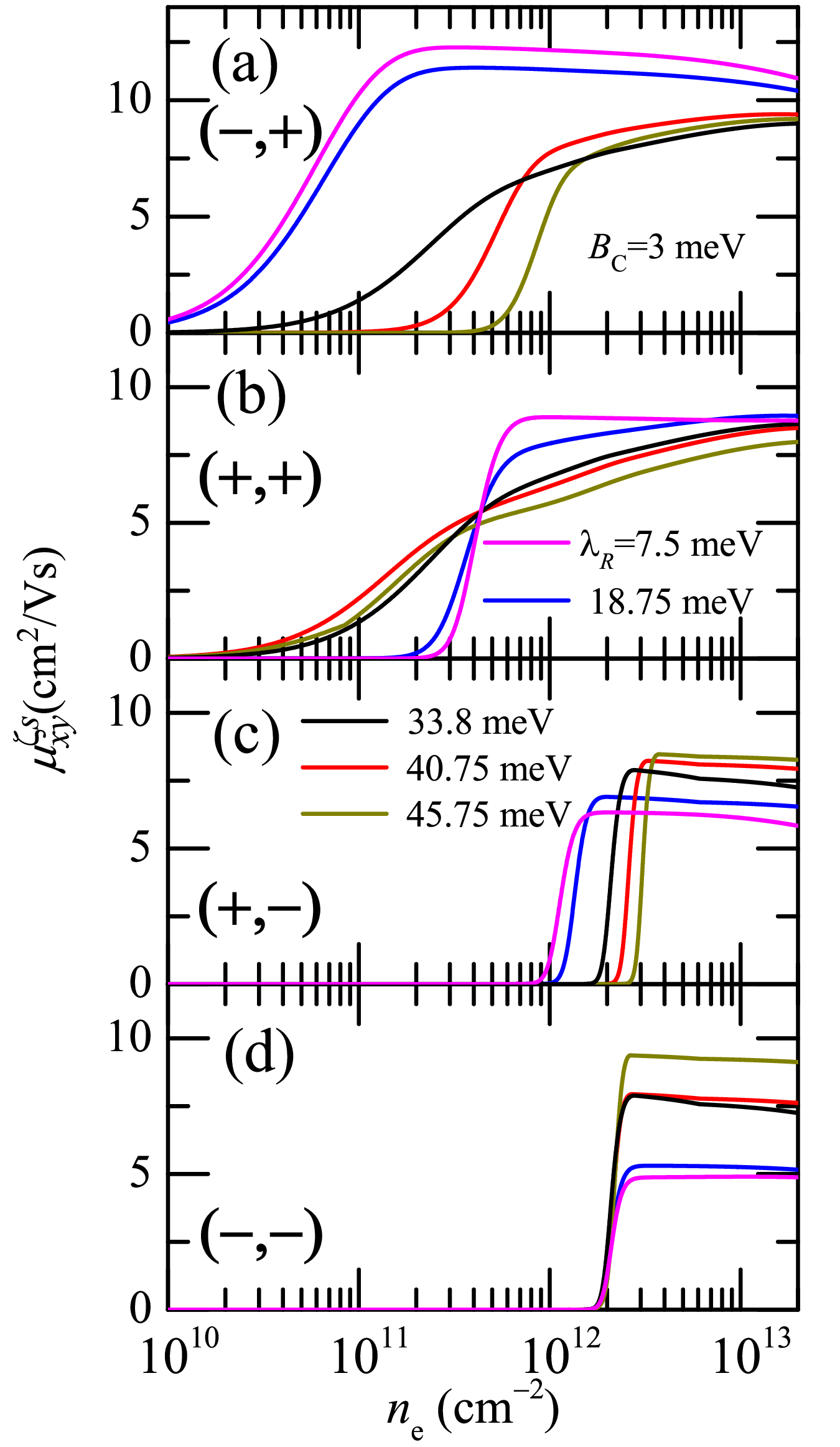}
\caption{The transverse or Hall mobility $\mu_{xy}^{\zeta s}$ in conduction
subband $(\zeta,s)=(\pm,\pm)$ in ML-MoS$_{2}$ as a function of electron
density at a fixed $B_c=3$ meV for different Rashba parameters $\lambda_{R}$
as indicated.}
\label{fig6}
\end{figure}

\begin{figure}[t]
\includegraphics[width=8.6cm]{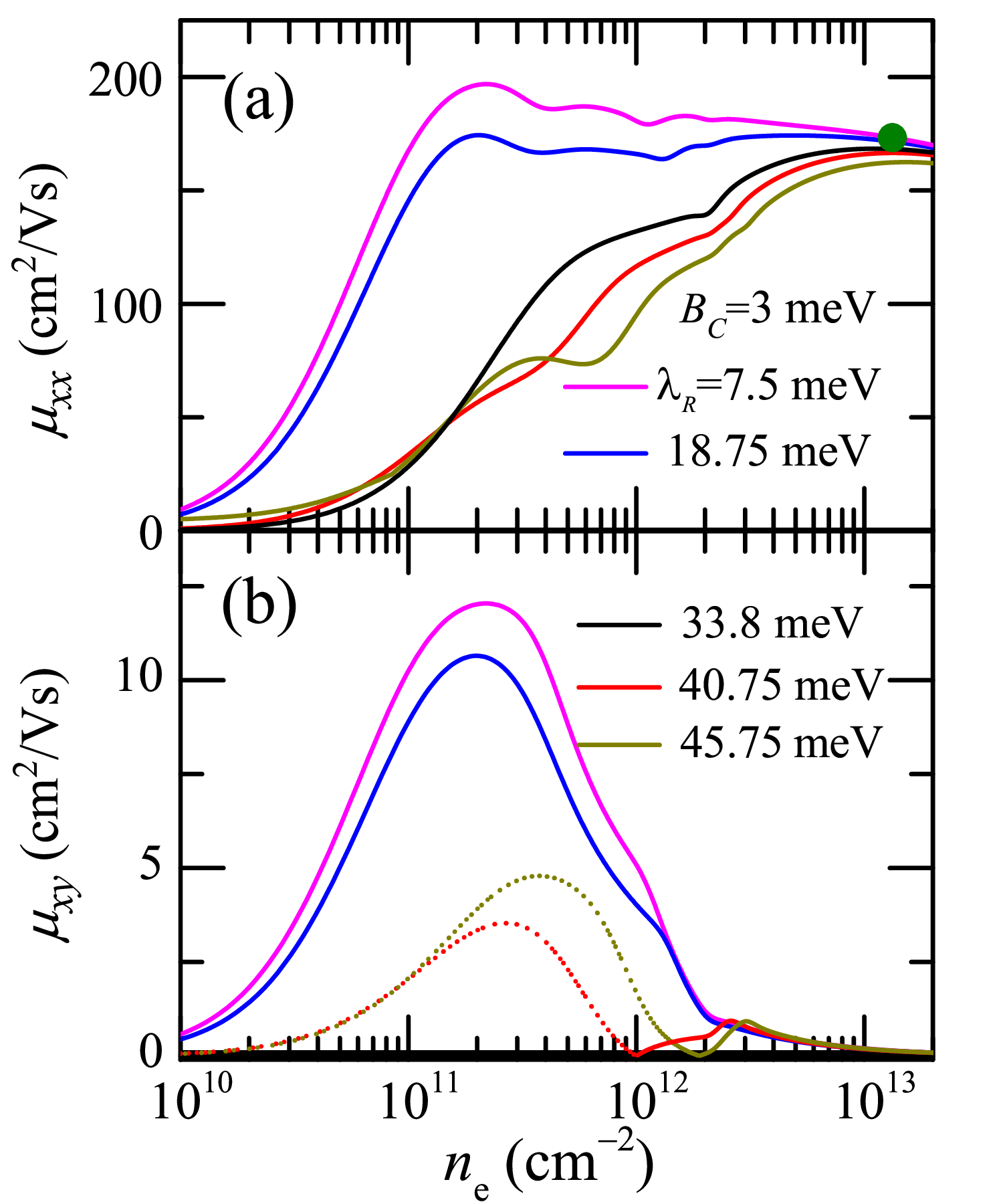}
\caption{(a) The longitudinal, $\mu_{xx}$, and (b) the transverse or Hall mobility,
$\mu_{xy}$, of ML-MoS$_{2}$ as a function of electron density at a fixed $B_c=3$ meV
for different Rashba parameters $\lambda_{R}$ as indicated. The green dot at
$n_e=1.35\times 10^{13}$ cm$^{-2}$ in (a) is the mobility obtained
experimentally \cite{Radisavljevic13}, from which we take the value of $N_i=3.98 \times10^{12}$
cm$^{-2}$ in our calculations. When $\lambda_R\approx 33.8$ meV, $\mu_{xy} \to 0$ in (b).
The dotted part of line corresponds to the reverse direction of the transverse current.}
\label{fig7}
\end{figure}

In the theoretical approach developed in this study for evaluating the mobilities,
the only fitting parameter that we need is the impurity concentration $N_i$ in
Eq. \eqref{transitionrate}. Here we take $N_i=3.98 \times 10^{12}$ cm$^{-2}$
for numerical calculations (see the green dot in \textcolor{blue}{Fig. \ref{fig7}(a))}. This value corresponds
to a longitudinal mobility obtained experimentally for an $n$-type ML-MoS$_2$
on a SiO$_2$/Si substrate \cite{Radisavljevic13}, which is about 174 cm$^2$V$^{-1}$s$^{-1}$
at $T=4$ K and $n_e=1.35 \times 10^{13}$ cm$^{-2}$. The experimental results
showed \cite{Radisavljevic13} that similar to semiconductor based 2D systems, the longitudinal
mobility of ML-MoS$_2$ depends weakly on temperature when $T< 10$ K. Thus, the
results obtained theoretically for $T\to 0$ from this study can be applied to
reproduce those obtained experimentally at low temperatures $T<10$ K for ML-MoS$_2$.
Furthermore, we take the transverse mobility in $K'$/$K$ valley is positive/negative
in Eq. \eqref{mobilityxy} due to different directions of the Hall-currents, akin to
the application of the magnetic fields along opposite directions.

In the presence of an external magnetic field, the Hall effect can lead to the transverse (Hall) mobility or conductivity \cite{Ando14}. In this study, the transverse mobility comes from the valley Hall effect driven by the pseudo-magnetic field induced by the Berry curvature. Namely the transverse current along the $y$-direction can be measured in the presence of a driving electric field applied along the $x$-direction owing to the lift of the valley degeneracy and to the achievement of the spin polarization in the electronic system \cite{Zhao20}. The Berry curvature in ML-TMDs can introduce a pseudo-magnetic field in each spin and valley subband. Thus, the nonzero total transverse (Hall) current or mobility can be observed since the valley Hall effects in two valleys can no longer be canceled out with each other in the presence of proximity-induced interactions . Furthermore, we note that in the presence of proximity-induced interactions, the electron energy is still symmetric along the $xy$-plane (see Eq. \eqref{diagonalized}), meaning that it depends only on $|k|$. However, the electron wavefunction is asymmetric in the $xy$-plane (see Fig. \eqref{eigenfunctions}), namely it depends on $k_x$ and $k_y$. Therefore, the integrations for off-diagonal elements in Eq. (34), which are related to Hall mobility $\mu_{xy}^{\zeta s}$, can be nonzero.

In Fig. \ref{fig5} and Fig. \ref{fig6}, we show respectively the longitudinal
$\mu_{xx}^{\zeta s}$ and transverse or Hall mobility $\mu_{xy}^{\zeta s}$
in conduction subband $(\zeta,s)=(\pm,\pm)$ in ML-MoS$_2$ as a function of
electron density $n_{e}$ at a fixed EZF $B_{c}=3$ meV for different Rashba
parameters $\lambda_{R}$. At $B_c=3$ meV, i) when $\lambda_R<33.8$ meV, the
conduction subbands from the lowest to the highest energies are $(-,+)$,
$(+,+)$, $(+,-)$ and $(-,-)$ (see Fig. \ref{fig2}(b)). Thus, $\mu_{xx}^{-+}$
and $\mu_{xy}^{-+}$ are always nonzero and $\mu_{xx}^{++}$ and $\mu_{xy}^{++}$,
$\mu_{xx}^{+-}$ and $\mu_{xy}^{+-}$ and $\mu_{xx}^{--}$ and $\mu_{xy}^{--}$
can be observed with increasing $n_e$ when they are occupied; ii) when $33.8 $
meV $<\lambda_R<59.9$ meV, the conduction subbands from the lowest to the highest
energies are $(+,+)$, $(-,+)$, $(-,-)$ and $(+,-)$ (see Fig. \ref{fig2}(b)).
Therefore, $\mu_{xx}^{++}$ and $\mu_{xy}^{++}$ are always nonzero and
$\mu_{xx}^{-+}$ and $\mu_{xy}^{-+}$, $\mu_{xx}^{--}$ and $\mu_{xy}^{--}$ and
$\mu_{xx}^{+-}$ and $\mu_{xy}^{+-}$ can be observed with increasing $n_e$ when
they become populated; and iii) when $\lambda_R\approx 33.8$ meV, the conduction
subbands from the lowest to the highest energies are $(+,+)\approx (-,+)$ and
$(-,-)\approx(+,-)$ (see Fig. \ref{fig2}(b)). As a result, $\mu_{xx}^{++}$,
$\mu_{xy}^{++}$, $\mu_{xx}^{-+}$ and $\mu_{xy}^{-+}$ are always nonzero and
$\mu_{xx}^{--}$, $\mu_{xy}^{--}$, $\mu_{xx}^{+-}$ and $\mu_{xy}^{+-}$ can be
observed with increasing $n_e$ when they are occupied. When a conduction subband
$(\zeta,s)$ becomes occupied, both $\mu_{xx}^{\zeta s}$ and $\mu_{xy}^{\zeta s}$
first increase rapidly then depend relatively weakly on $n_e$ with increasing
$n_e$. Moreover, we notice that the longitudinal and transverse mobilities
show almost the identical dependence upon the electron density. In the magneto-transport measurements
in the presence of an external magnetic field,
the longitudinal (drift) and transverse (Hall) mobilities in a 2D electron gas have a relationship \cite{Ando14}:
$\mu_{xy}=r_\mathrm{H}\mu_{xx}=\sigma R_\mathrm{H}$, where $r_\mathrm{H}$ is the Hall scattering factor which
is a constant for a material or device and $R_\mathrm{H}$ is the Hall coefficient. In
the present study, the valley Hall effect occurs with the presence of the pseudo-magnetic
field induced by the Berry curvature in each spin and valley split subband in a valley nondegenerate
ML-TMDs system. Therefore, the contributions to the longitudinal and transverse mobilities
from each subband also have the similar features with a proportional relationship.
Furthermore, $\mu_{xx}^{\zeta s}$ is about 20 times larger than
$\mu_{xy}^{\zeta s}$ when $(\zeta,s)$ is well occupied. These results indicate that when $B_c \neq 0$, $\mu_{xx}^{\zeta s}$ and $\mu_{xy}^{\zeta s}$ in an
$n$-type ML-MoS$_2$ depend strongly on Rashba parameter and electron density.

\begin{figure}[t]
\includegraphics[width=8.6cm]{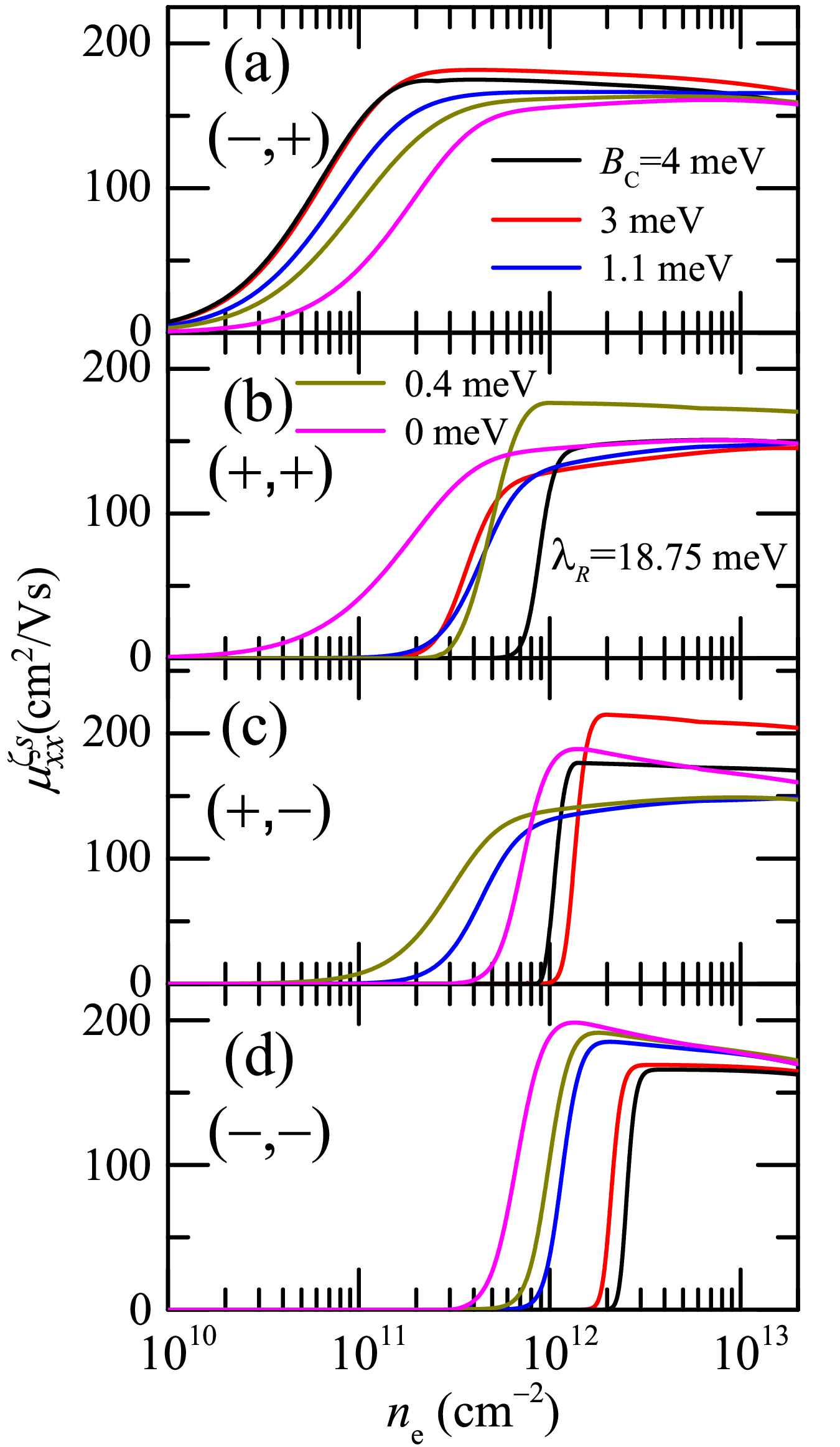}
\caption{The longitudinal mobility $\mu_{xx}^{\zeta s}$ in conduction subband
$(\zeta,s)=(\pm,\pm)$ in ML-MoS$_{2}$ as a function of electron density at a fixed
Rashba parameters $\lambda_{R}=18.75$ meV for different EFZ $B_c$ as indicated.}
\label{fig8}
\end{figure}

In Fig. \ref{fig7}, we show the averaged longitudinal $\mu_{xx}$ and transverse
or Hall mobility $\mu_{xy}$ in $n$-type ML-MoS$_2$ as a function of electron density
$n_{e}$ at a fixed EZF $B_{c}=3$ meV for different Rashba parameters $\lambda_{R}$,
obtained by using Eqs. \eqref{mobilityxx}-\eqref{mobilityxy}.
In low $n_e$ regime $\mu_{xx}$ increases with $n_e$.
In high $n_e$ regime, $\mu_{xx}$ depends relatively weakly on $n_e$ and $\mu_{xx}$
decreases with increasing $\lambda_R$. We find that when $B_c\neq 0$, $\mu_{xy}\neq 0$
can be observed and the features of $\mu_{xy}$ differ significantly from those of
$\mu_{xx}$. (i) $\mu_{xy}$ is about 20 times smaller than $\mu_{xx}$; (ii) $\mu_{xy}$
first increases then decreases with increasing $n_e$. This is due to the fact that
at low $n_e$ only the lowest conduction subband is occupied. With increasing $n_e$
and when the higher subbands with different valleys index becomes populated, the
valley-currents from different subbands are offset partly and, thus, the overall
$\mu_{xy}$ decreases with increasing $n_e$. With further increasing $n_e$ and when
all four conduction subbands become occupied, the overall $\mu_{xy}$ further decreases
and approaches to zero gradually. These results suggest that $\mu_{xy}\neq 0$ can be
observed in relatively low $n_e$ in $n$-type ML-MoS$_2$. (iii) When $\lambda_{R}\approx33.8$ meV,
$\mu_{xy} \to 0$ because $(-,+)\approx (+,+)$ and $(-,-)\approx(+,-)$ so that
$n_+^+ \mu_{xy}^{++} -n_+^- \mu_{xy}^{-+}\to 0$ and $n_-^- \mu_{xy}^{--} -n_-^+ \mu_{xy}^{+-}\to 0$;
And (iv) interestingly, the sign of $\mu_{xy}$ changes at about $\lambda_{R}=33.8$ meV
and the sign of $\mu_{xy}$ would change again after approached to zero with increasing carrier
density when $\lambda_R=40.75$ meV and $\lambda_R=45.75$ meV [see in Fig. \ref{fig7}(b)],
implying that the direction of
the Hall current/voltage can be varied through electrically tuning of the Rashba parameter
and carrier density through, e.g.,  the presence of a substrate and/or tuning the
applied gate voltage.
The transverse mobility has often been investigated in 2D electron gas systems through the Hall effect in magneto-transport measurement \cite{Sucharitakul15}. The experimental investigation
of the transverse mobility can be conducted in a field effect transistor or encapsulated
ML-MoS$_2$ multi-terminal devices \cite{Radisavljevic13,Cui15}.

\begin{figure}[t]
\includegraphics[width=8.6cm]{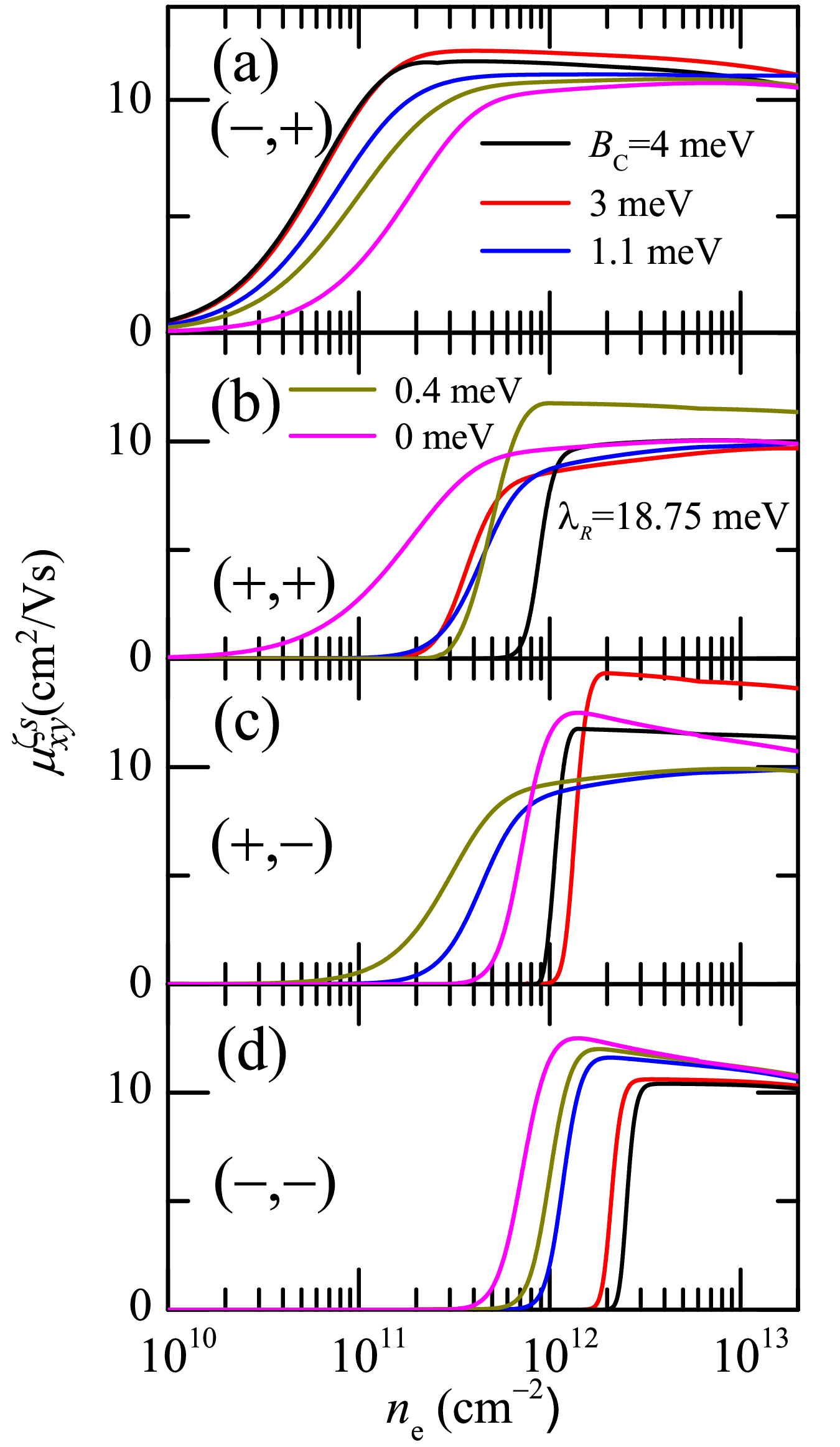}
\caption{The transverse or Hall mobility $\mu_{xy}^{\zeta s}$ in conduction
subband $(\zeta,s)=(\pm,\pm)$ in ML-MoS$_{2}$ as a function of electron density
at a fixed Rashba parameters $\lambda_{R}=18.75$ meV for different EZF $B_c$
as indicated.}
\label{fig9}
\end{figure}

\begin{figure}[t]
\includegraphics[width=8.6cm]{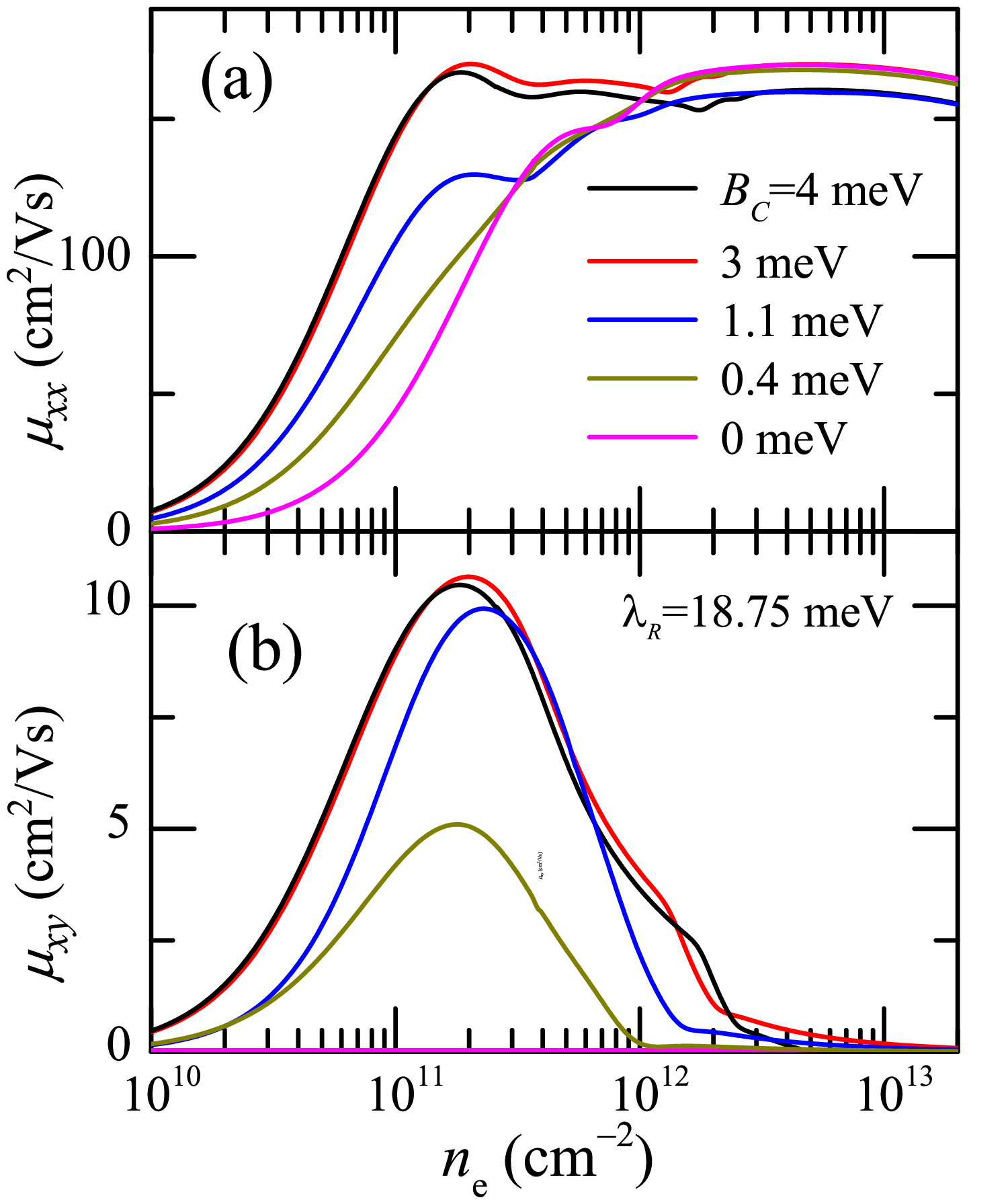}
\caption{(a) The transport longitudinal mobility $\mu_{xx}$ and (b)
the transport transverse or Hall mobility $\mu_{xy}$ of ML-MoS$_{2}$
as a function of electron density at a fixed Rashba parameter $\lambda_{R}=18.75$ meV for
with different EZF $B_c$. In (b), $\mu_{xy}=0$ when $B_c=0$.}
\label{fig10}
\end{figure}

In Fig. \ref{fig8} and Fig. \ref{fig9}, we show respectively the longitudinal
$\mu_{xx}^{\zeta s}$ and transverse or Hall mobility $\mu_{xy}^{\zeta s}$ in
conduction subband $(\zeta,s)=(\pm,\pm)$ in ML-MoS$_2$ as a function of electron
density $n_{e}$ at a fixed Rashba parameter $\lambda_R=18.75$ meV for different
EZF $B_c$. At $\lambda_R=18.75$ meV (see Fig. \ref{fig2}(a)), i) the lowest conduction
subband is always $(-,+)$ so that $\mu_{xx}^{-+}$ and $\mu_{xy}^{-+}$ are always
nonzero; ii) when $B_c<1.1$ meV, the conduction subbands from the lowest to the
highest energies are $(-,+)$, $(+,-)$, $(+,+)$ and $(-,-)$ (see Fig. \ref{fig2}(a)).
Thus, $\mu_{xx}^{-+}$ and $\mu_{xy}^{-+}$ are always nonzero and $\mu_{xx}^{+-}$
and $\mu_{xy}^{+-}$, $\mu_{xx}^{++}$ and $\mu_{xy}^{++}$ and $\mu_{xx}^{--}$ and
$\mu_{xy}^{--}$ can be observed with increasing $n_e$ when they are occupied; ii)
when $1.1 $ meV $<B_c<4$ meV, the conduction subbands from the lowest to the
highest energies are $(-,+)$, $(+,+)$, $(+,-)$ and $(-,-)$ (see Fig. \ref{fig2}(a)).
Therefore, $\mu_{xx}^{-+}$ and $\mu_{xy}^{-+}$ are always nonzero and $\mu_{xx}^{++}$
and $\mu_{xy}^{++}$, $\mu_{xx}^{+-}$ and $\mu_{xy}^{+-}$ and $\mu_{xx}^{--}$ and
$\mu_{xy}^{--}$ can be observed with increasing $n_e$ when they become populated;
and iii) when $B_c\approx 1.1$ meV, the conduction subbands from the lowest to the
highest energies are $(+,+)\approx (+,-)$ (see Fig. \ref{fig2}(a)). As a result, $\mu_{xx}^{-+}$
and $\mu_{xy}^{-+}$ are always nonzero and $\mu_{xx}^{++}$, $\mu_{xy}^{++}$,
$\mu_{xx}^{+-}$, $\mu_{xy}^{+-}$, $\mu_{xx}^{--}$ and $\mu_{xy}^{--}$ can be observed
with increasing $n_e$ when they are occupied. When a conduction subband $(\zeta,s)$
becomes occupied, both $\mu_{xx}^{\zeta s}$ and $\mu_{xy}^{\zeta s}$ first increase
rapidly then depend relatively weakly on $n_e$ with increasing $n_e$. Furthermore,
$\mu_{xx}^{\zeta s}$ is about 20 times larger than $\mu_{xy}^{\zeta s}$ when $(\zeta,s)$
is well occupied. These results indicate that when $\lambda_R \neq 0$, $\mu_{xx}^{\zeta s}$
and $\mu_{xy}^{\zeta s}$ in an $n$-type ML-MoS$_2$ depend strongly on EZF parameter
and electron density.

In Fig. \ref{fig10}, we show the averaged longitudinal $\mu_{xx}$ and
transverse or Hall mobility $\mu_{xy}$ in $n$-type ML-MoS$_2$ as a function
of electron density $n_{e}$ at a fixed Rashba parameter $\lambda_{R}=18.75$
meV for different EZF parameters $B_c$, obtained by using
Eqs. \eqref{mobilityxx} and \eqref{mobilityxy}. In low $n_e$ regime $\mu_{xx}$
increases with $n_e$. In high $n_e$ regime, $\mu_{xx}$ depends relatively
weakly on $n_e$ and $\mu_{xx}$ decreases with increasing $B_c$. We find that
when $B_c\neq 0$, $\mu_{xy}\neq 0$ can be observed and the features of
$\mu_{xy}$ differ significantly from those of $\mu_{xx}$. (i) $\mu_{xy}$ is
about 20 times smaller than $\mu_{xx}$; (ii) $\mu_{xy}$ first increases then
decreases with increasing $n_e$. This is due to the fact that at low $n_e$
only the lowest conduction subband is occupied. With increasing $n_e$ and
when the higher subbands with different valleys index becomes populated,
the valley-currents from different subbands are offset partly and, thus,
the overall $\mu_{xy}$ decreases with increasing $n_e$. With further
increasing $n_e$ and when all four conduction subbands become occupied,
the overall $\mu_{xy}$ further decreases and approaches to zero gradually.
These results suggest that $\mu_{xy}\neq 0$ can be observed in relatively
low $n_e$ in $n$-type ML-MoS$_2$. (iii) When $B_c\approx 0$ meV, $\mu_{xy} \to 0$
because the energies of the bands in different valley are degenerate where
$(+,-)=(-,+)$ and $(-,-)=(+,+)$ so that $n_-^+ \mu_{xy}^{+-}-n_+^-\mu_{xy}^{-+}\to 0$
and $n_-^- \mu_{xy}^{--}-n_+^+ \mu_{xy}^{++}\to 0$. These results indicate
that the longitudinal and transverse mobilities can also been tuned by
the EZF parameter and carrier density. Comparing
Fig. \ref{fig7} with Fig. \ref{fig10}, we see that both of the Rashba
parameter and the EZF parameter $B_c$ can effectively modulated the longitudinal
and transverse mobilities. The stronger effect by varying $B_c$ can be observed.
However, a large Rashba SOC can reverse the direction of the transverse current
with varying electron density.
With a relatively large $n_e$, the effect of the Rashba
parameter and the EZF parameter on longitudinal
and transverse mobilities becomes weak due to the weakening of the spin polarization in the sample system \cite{Zhao20}. When $n_e$ is large enough so that all spin and valley subbands are occupied, $\mu_{xy}$ begins to vanish.

\begin{figure}[t]
\includegraphics[width=8.6cm]{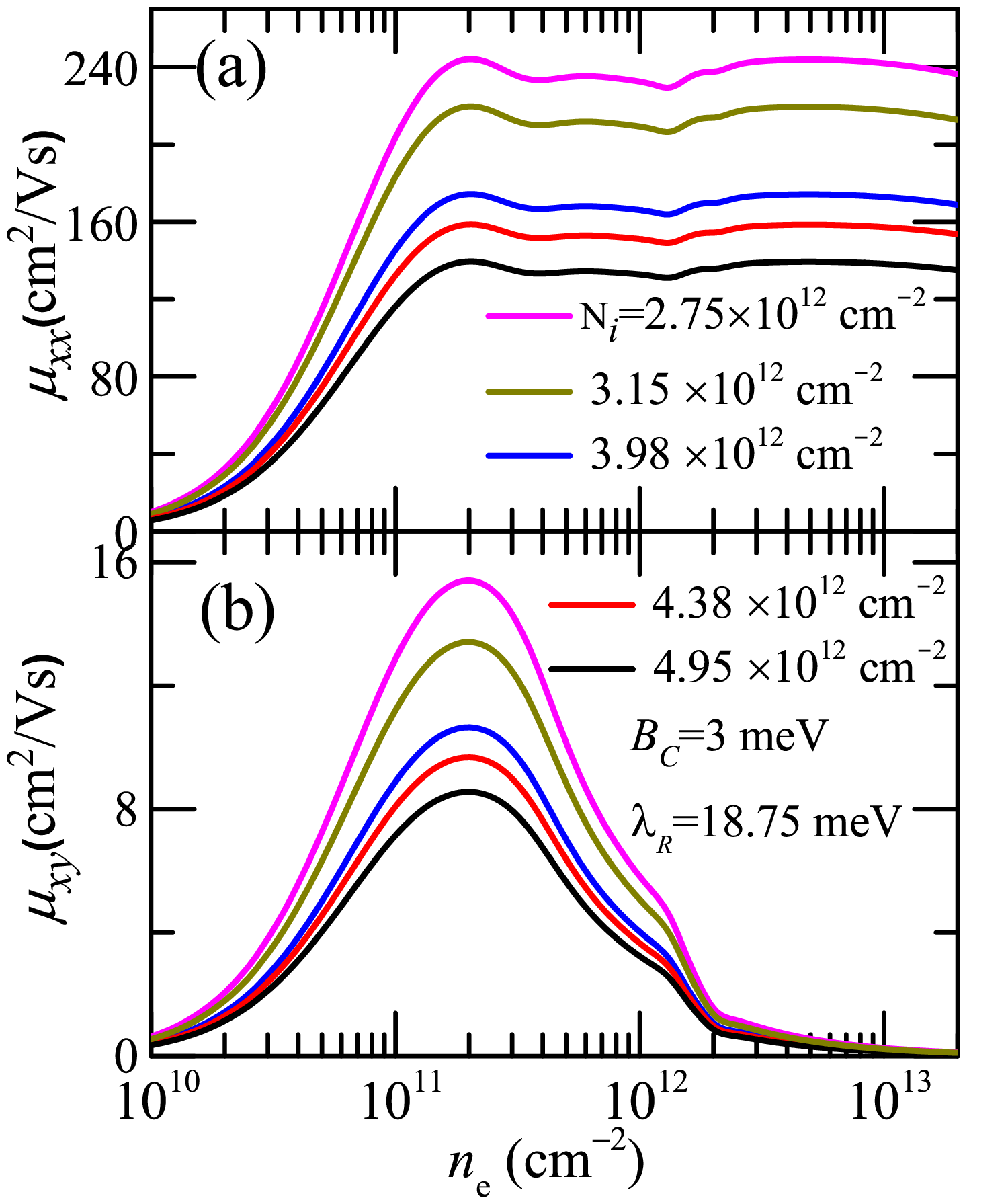}
\caption{(a) The transport longitudinal mobility $\mu_{xx}$ and (b)
the transport transverse or Hall mobility $\mu_{xy}$ of ML-MoS$_{2}$
as a function of electron density at a fixed Rashba parameter
$\lambda_{R}=18.75$ meV and EZF parameters $B_c$=3 meV and $B_v$= 5 meV for different
impurity concentrations $N_i$.}
\label{fig11}
\end{figure}

In Fig. \ref{fig11}, we plot the averaged longitudinal $\mu_{xx}$ and
transverse $\mu_{xy}$ mobility in $n$-type ML-MoS$_2$ as a function
of electron density $n_{e}$ at a fixed Rashba parameter $\lambda_{R}=18.75$
meV, and EZF parameters $B_c$ and $B_v$ for different impurity concentrations $N_i$. As we can see,
both $\mu_{xx}$ and $\mu_{xy}$ decrease with increasing $N_i$ because a larger $N_i$ corresponds to a stronger electron-impurity scattering rate.
In our theoretical model developed in this study, the transverse mobility is attributed
to both the intrinsic contribution by Berry curvature and the extrinsic
contribution by impurity scattering \cite{Nagaosa10,Xiao10}. With a relatively large impurity concentration
and at low temperatures, the impurity scattering is the principal channel for electronic scattering rather than phonon
scattering. The impurity scattering refers to the skew
scattering mechanism, which is proportional to the momentum relaxation time. Since the
impurity scattering rate can significantly affect the electron lifetime, the
extrinsic impurity scattering would play an important role to affect both the
longitudinal and transverse mobilities, as shown in Fig. \ref{fig11}.
As far as we know, there is a lack of experimental work for measuring the Hall mobility in ML-TMDs in
the absence of an external magnetic field when the valley degeneracy is lifted
by proximity-induced interactions, as we predict here. The method of the
measurement on such an effect should be the same as that applied for conventional Hall measurement in magneto-transport experiments by using the Hall bar or van der Pauw electrodes.

In this study, we considered an $n$-type ML-MoS$_2$ laid on a ferromagnetic
substrate where the proximity-induced interactions are presented. The electronic
screening and transport mobility are contributed by the electrons in the
spin-splitting subbands. The electronic structure of an $n$-type ML-MoS$_2$/ferromagnetic
substrate heterostructure can be effectively tuned by varying the Rashba SOC
and EZF via the types of substrate or a perpendicular electrical field.
As we know, the carrier density in a 2D system can also be effectively
tuned through, e.g., applying a gate voltage. Thus, the electronic
screening and transport mobility of a ML-MoS$_2$ heterostructure
system can be tuned by varying these parameters through the state-of-the
-art fabrication method of devices and experimental settings. More interestingly,
there exists transverse or Hall mobility due to the breaking of valley
degeneracy by the EZF. The longitudinal and transverse or Hall mobility
can be effectively tuned by the Rashba parameter, EZF, and carrier
density. With the unique proximity-induced interactions, a ML-MoS$_2$-based heterostructure can be a promising material for electronics and
valleytronics.

\section{Conclusions}
\label{sec:conclusions}

In this paper, we theoretically investigate the electronic and transport
mobility properties of an $n$-type ML-MoS$_{2}$ at low temperature in the
presence of proximity-induced interactions such as Rashba SOC and exchange
interaction. The electric screening induced by electron-electron interaction
is studied under a standard RPA, and the longitudinal and transverse or Hall
mobilities are evaluated by using a momentum-balance equation derived
from a semi-classical Boltzmann equation where the electron-impurity
interaction is considered as the principal scattering event at low temperature.
We have examined the roles of Rashba SOC, EZF, and carrier density on affecting
the occupation of electrons in spin splitting subbands in different valleys, the
inverse screening length, and longitudinal and transverse or Hall mobility of an
$n$-type ML-MoS$_{2}$. The main conclusions obtained from this study are
summarized as follows.

In a 2D ML-TMDs material such as ML-MoS$_{2}$, the Rashba SOC can result in an
in-plane electronic spin component. The presence of the proximity-induced
exchange interaction can further modify the spin splitting and lift the
energy degeneracy for electrons in different valleys. The opposite signs
of Berry curvatures in the two valleys would introduce opposite directions
of Lorentz force on valley electrons. The inverse screening lengths and
transport longitudinal and transverse or Hall mobilities are different in each
spin splitting subbands due to the electronic structure and the occupations
of electrons in each spin splitting subbands. The total mobility
is contributed by each spin splitting subbands. Due to the breaking of valley degeneracy by
the EZF, the currents from different valleys are no longer canceled out so
that the transverse current or Hall mobility can be observed in the absence
of an external magnetic field. The electronic screening, longitudinal mobility,
and transverse or Hall mobility can be effectively tuned by the electron density and
proximity-induced interactions with the Rashba effect and exchange interaction with
an effective Zeeman field. We find that at a fixed effective
Zeeman field, the lowest spin-split conduction subband in ML-TMDs can be tuned
from one in the $K'$-valley to one in the $K$-valley by varying the Rashba parameter.
Therefore, we can change the magnitude and direction of the Hall
current by varying the Rashba parameter and/or the effective Zeeman field
regarding the proximity effect induced by, e.g., the presence of a ferromagnetic
substrate and/or applying a gate voltage. As the Hamiltonian
for different ML-TMDs is the same but with different material parameters, the
behaviors of the longitudinal and transverse (Hall) mobilities
should show similar features for different ML-TMD systems in the presence of
proximity-induced interactions.
The important and interesting theoretical findings in this paper can
be beneficial to experimental observation of the valleytronic effect and
to gaining an in-depth understanding of the ML-TMDs systems in the presence
of proximity-induced interactions. We hope that the theoretically predictions
in this work can be verified experimental in the near future.

\section*{ACKNOWLEDGMENTS}
This work was supported by the National Natural Science foundation
of China (NSFC) (Grants  No. U2230122, No. U2067207, No. 12364009, and No. 12004331),
Shenzhen Science and Technology Program (Grant
No. KQTD20190929173954826), and by Yunnan Fundamental Research
Projects (Grants No. 202301AT070120, and No. 202101AT070166), Y.M.X was supported through the
Xingdian Talent Plans for Young Talents of Yunnan Province (Grant
No. XDYC-QNRC-2022-0492).

\end{document}